\documentclass[aps,amsmath,amssymb,reprint]{revtex4-1}
\pdfoutput=1
\usepackage{graphicx}
\usepackage{dcolumn}
\usepackage{bm}
\usepackage{verbatim}

\begin{document}

\preprint{May 9, 2014}

\title[Multi-state iterative Boltzmann Inversion]{
Derivation of coarse-grained potentials via multistate iterative Boltzmann inversion}

\author{Timothy C. Moore}
\author{Christopher R. Iacovella}%
\affiliation{Department of Chemical and Biomolecular Engineering, Vanderbilt University,
Nashville,~TN~37235~USA}
\affiliation{Vanderbilt Multiscale Modeling and Simulation (MuMS) Center, Vanderbilt
University, Nashville,~TN~37235~USA}
\author{Clare McCabe} \email{c.mccabe@vanderbilt.edu}
\affiliation{Department of Chemical and Biomolecular Engineering, Vanderbilt University,
Nashville,~TN~37235~USA}
\affiliation{Vanderbilt Multiscale Modeling and Simulation (MuMS) Center, Vanderbilt
University, Nashville,~TN~37235~USA}
\affiliation{Department of Chemistry, Vanderbilt University,
Nashville,~TN~37235~USA}
\date{\today}

\begin{abstract}
  In this work, an extension to the standard iterative Boltzmann inversion (IBI) method to 
  derive coarse-grained potentials is proposed. 
  It is shown that the inclusion of target data from multiple states yields a less
  state-dependent potential, and is thus better suited to simulate systems over a range of
  thermodynamic states than the standard IBI method.
  The inclusion of target data from multiple states forces the algorithm to sample regions of 
  potential phase space that match the radial distribution function at multiple state
  points, thus producing a derived potential that is more representative of the underlying 
  potential interactions. 
  It is shown that the algorithm is able to converge to the true potential for a system
  where the underlying potential is known. It is also shown that potentials derived via the
  proposed method better predict the behavior of \textit{n}-alkane chains
  than those derived via the standard method. 
  Additionally, through the examination of alkane monolayers, it is shown that the relative
  weight given to each state in the fitting procedure can impact bulk system properties,
  allowing the potentials to be further tuned in order to match the properties of reference
  atomistic and/or experimental systems.
\end{abstract}

\keywords{coarse-grained, molecular simulation,
forcefields, iterative Boltzmann inversion}

\maketitle
\section{\label{sec:intro}Introduction}
The utility of coarse-grained (CG) forcefields for soft matter and biological simulations has been well established in the literature, enabling simulation to explore greater length- and time-scales than is feasible with fully atomistic models. 
This is of particular importance when studying the self-assembly of soft matter systems, where the
assembly is typically driven by weak forces, (e.g., hydrophobicity and entropy)
\cite{israelachvili1976theory,israelachvili1977theory,muvsevivc2008self,bishop2009nanoscale,iacovella2011self,ziherl2001maximizing}
 and structures often demonstrate hierarchical ordering (e.g., molecules organized into micelles,
 micelles organized into local/global patterns).
 \cite{iacovella2011self,thomas1987ordered,phillips2010stability,iacovella2009complex,glaser2007soft,ziherl2000soap}
While generic, non-specific CG models have been widely applied,
\cite{larson1985monte,glotzer2005self,gai2012wang,iacovella2005phase,lau1989lattice}
providing important information regarding trends and design rules, it is often necessary to use CG models specifically mapped to the system of interest to provide a direct one-to-one correspondence with experiment. 
While several “transferable” CG forcefields, such as TraPPE-CG
\cite{maerzke2011transferable} and MARTINI,\cite{marrink2007martini} have been developed, akin to
forcefield development at the atomistic level,
\cite{martin1998transferable,jorgensen1996development,oostenbrink2004biomolecular,vanommeslaeghe2010charmm,wang2004development}
the development of new CG forcefields is still often necessary.
This is often required since the available forcefields may be lacking the necessary molecular
species/groupings or may not have been optimized for the properties of interest.
This second point is of particular consequence, since, for example, a forcefield optimized to match
phase behavior may not appropriately capture subtle structural
features.\cite{maerzke2011transferable}
Generally speaking, direct structural correspondence is needed to accurately transition between
different simulation levels (e.g., atomistic to CG), in order to perform multiscale
\cite{mccabe2004multiscale,liu2011multiscale,peter2009multiscale} and hybrid-multiscale simulations,
\cite{di2012mixing,ensing2007energy,lidorikis2001coupling,michel2008prediction,praprotnik2008multiscale,rzepiela2011hybrid,werder2005hybrid}
as well as to recover atomistic details from CG simulations.
\cite{heath2007coarse,hess2006osmotic,liu2008reconstructing,rzepiela2010reconstruction} 

Several approaches have been developed to derive and optimize CG 
forcefields.
\cite{reith2003deriving,ercolessi1994interatomic,izvekov2005multiscale,shell2008relative,chaimovich2011coarse,iacovella2010pair,bozorgui2013fluctuation} 
Among these, the iterative Boltzmann inversion (IBI) method\cite{reith2003deriving} has 
become a popular choice due to its straightforward nature, general applicability to a wide 
range of systems, and basis in structural properties.
The IBI method relies on self-consistently adjusting a given potential to achieve
convergence with target structural 
data; for nonbonded interactions this target data takes the form of the radial distribution 
function (RDF) between interaction sites and the potential is iteratively updated 
according to:
\begin{equation}
  V_{i+1}(r) = V_i(r) + \alpha k_B T \ln\left[\frac{g^i(r)}{g^*(r)} \right]
  \label{eq:ss-ibi}
\end{equation}
where $V_i(r)$ is a numerical pair potential; $i$ represents the current iteration;
$\alpha$ is a damping factor to suppress larges changes to the potential update, often
varying from 0.2 to unity, where smaller values tend to be necessary to capture dense
and/or crystalline states;\cite{hadley2010coarse}
$k_B$ is the Boltzmann constant; $T$ the absolute temperature; 
$r$ the separation between particles; 
$g_i(r)$ the pair RDF from the simulation of $V_i(r)$,
and $g^*(r)$ the RDF of the target system mapped to the CG level.
Although the CG potentials derived from IBI are typically able to accurately reproduce the
target RDFs, they are, in general, only applicable at the state point for which they were
derived, due to the structural nature of their derivation (e.g., note the explicit
temperature dependence of Equation 1, as well as the implicit temperature and density
dependence through the $g(r)$ terms).\cite{faller2004automatic,fu2012test}
For example, separate potentials were required to capture both the solid and fluid
structures of a pure simple lipid;\cite{hadley2010coarse}
Qian, \textit{et al.}\cite{qian2008temperature} found that the potential derived using IBI
for ethylbenzene scales in a non-linear fashion with temperature (i.e., a square root
dependence); 
and several works have shown that CG polymer potentials derived via the IBI method can depend
on the chemical environment for which they were derived.
\cite{reith2003deriving,bayramoglu2012coarse,bayramoglu2013modeling}
Recent work has shown that some of the CG potentials in a benzene-urea-water system derived
via IBI have some degree of state point transferability,\cite{ganguly2013convergence} but
it is unclear why IBI provides transferability for some but not all.
Furthermore, for complex systems, it may not be possible to optimize the potentials at the
state points of interest, due to time- or length-scale limitations of the atomistic
simulations and/or system complexity (i.e., systems many unique interactions that need to
be derived simultaneously); thus making it difficult to apply the IBI method appropriately,
given that potentials are not necessarily transferable.
Perhaps of most concern is the fact that the IBI method does not guarantee a unique
solution, as a multitude of vastly differing potentials may give rise to otherwise
matching RDFs. 
The form of the final derived potential may also vary based on runtime parameters, such as
the inital potential guess, potential cutoff, magnitude of the damping factor, etc.
Additionally, the derived potential may include artifacts associated with intermediate and
long-range structural correlations in the system, e.g., oscillatory behavior in the
potential that follows the peaks and valleys in the RDF, which may alter other properties
of the system, even if RDFs match.

\begin{figure}[h]
  \centering
  \includegraphics[width=5.334cm,keepaspectratio]{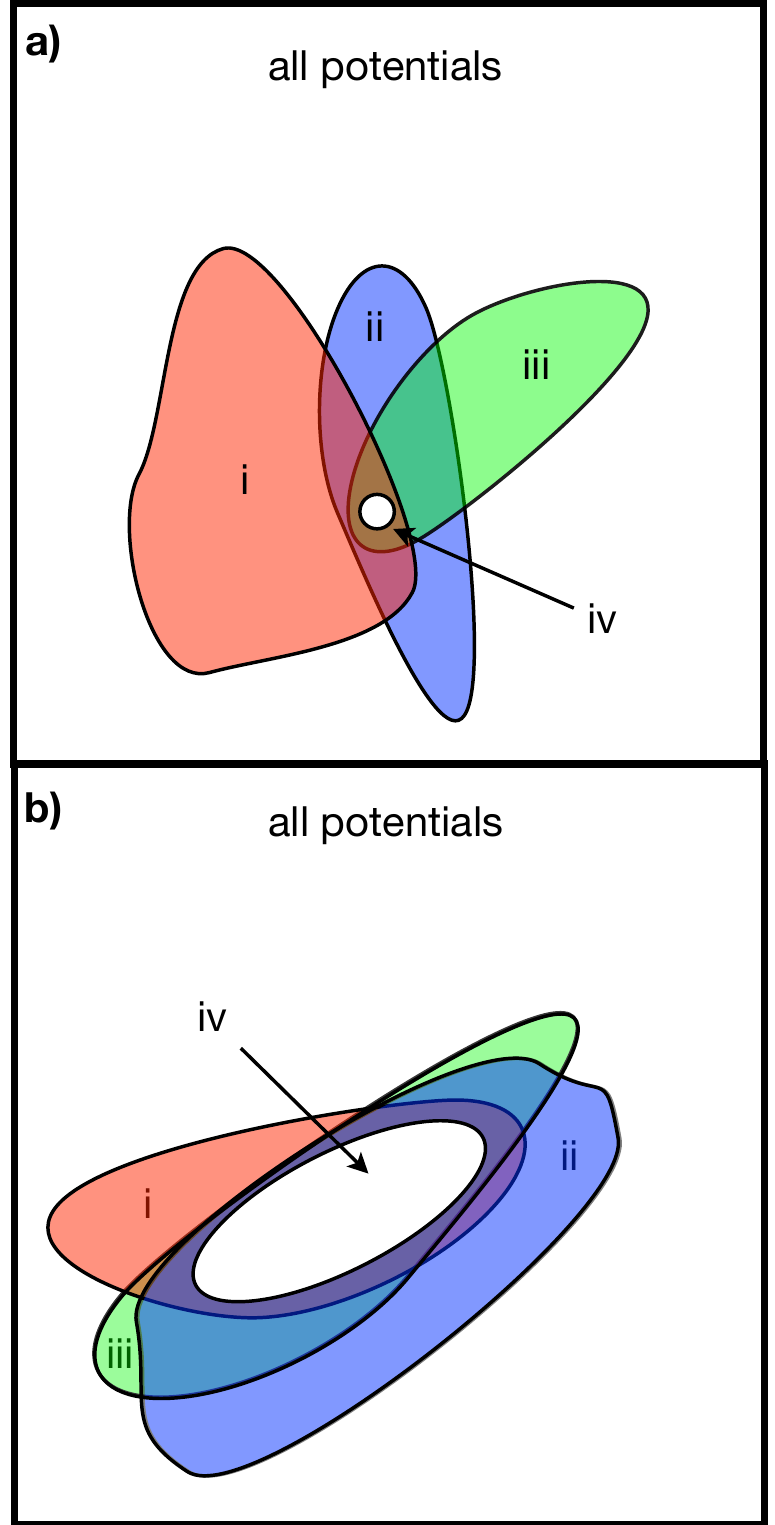}
  \caption{Regions of good potential phase space for states with optimal overlap (top) and
  too much overlap (bottom).}
  \label{fig:pot-space}
\end{figure}

In this work, the IBI method is extended to perform multi-state optimization, i.e., the
potential is self-consistently adjusted to achieve simultaneous convergence of target data
from multiple states.
The general idea, illustrated in Figure~\ref{fig:pot-space}, is that the inclusion of
target data from multiple states adds constraints to the optimization problem, such that
the derived forcefield tends toward a single potential that can adequately represent all
states.
For example, potentials in region `$i$' of the upper portion of Figure~\ref{fig:pot-space}
are able to match the target structure at a single state `$i$', potentials in region `$ii$'
are able to reproduce target data at state `$ii$', etc., with a single representative
potential lying at the overlap of these regions, shown as region `$iv$'. 
To test the efficacy of the proposed multi-state iterative Boltzmann inversion (MS
IBI) method, in Section~\ref{subsec:mlj}, we first perform potential optimizations for the
idealized system of a Lennard-Jones (LJ) fluid for which the potential is known, in order
to determine if the method resolves the correct potential.
In Section~\ref{subsec:propane}, to test the method in a system where only nonbonded
interactions are present in the CG model, a 3-to-1 mapped CG forcefield is optimizted for
propane using target data generated from united-atom (UA) propane simulations, and compared
with a single-site LJ model mapped to the experimental critical point of propane.
In Section~\ref{subsec:dodecane}, we apply this approach to \textit{n}-dodecane, a system
more representative of the typical application of a CG forcefield.
In Section~\ref{subsec:monolayers}, we examine a monolayer system composed of
\textit{n}-docecane, where it is demonstrated that adjustment of the relative weights given
to each target in the MS IBI method can be used to tune the potentials to match other
measurable system properties beyond the RDF.

\section{\label{sec:methods}Methods and Simulation Details}
\subsection{\label{subsec:ss-ibi}Single-state iterative Boltzmann inversion}
In the IBI method (which for clarity we shall refer to as single state, SS IBI), a
numerical pair potential, $V(r)$ is iteratively updated according to
Equation~\ref{eq:ss-ibi}.
In this manner, $V(r)$ is updated at each separation, $r$, based on whether the RDF
overpredicts or underpredicts the target RDF at the given $r$, and is repeated until the
trial RDF matches the target RDF within some tolerance.\cite{reith2003deriving}
The initial guess of the numerical potential is often taken to be the Boltzmann inversion
of the RDF of the target system:
\begin{equation}
  V_0(r) = -k_B T \ln g^*(r)
  \label{eq:ss-pmf}
\end{equation}
While not exact for site-site interactions in molecules,\cite{chan2007coarse} this
methodology is motivated by the statistical mechanics relationship between the potential of
mean force (PMF) and the RDF, and provides a reasonable starting potential over which to
iterate.

Typically, potentials derived with this method are capable of reproducing the target RDFs
with high accuracy, with slight deviations resulting from information lost during
coarse-graining.
The ease of use of the IBI method and its general applicability make it a powerful
tool; given a CG mapping and a target RDF, site-site pair potentials can be readily derived
with little user input.

\subsection{\label{subsec:ms-ibi}Multistate extension of IBI}
Although potentials derived with SS IBI will typically reproduce their target RDFs with
high accuracy, caution must be taken when using the potentials.
Upon successful convergence of the potential, it is only guaranteed that the derived and
target RDFs match, not that the potential is necessarily representative of the ``true''
underlying potential (i.e, not necessarily state independent).
It is important to note, especially since information is lost due to coarse-graining, that
a multitude of potentials may give rise to similar RDFs.
Only a small portion of the potentials that produce matching RDFs may actually fall within
the region of potentials that match the true potential and, since the true potential is
typically unknown, it is difficult to assess the accuracy of the derived potentials.
If the derived potential falls far outside the true potential region, this may give rise to
potentials that, despite providing a good match for the target RDF, lack transferability
and may contain artifacts making them incapable of resolving system properties other than
the RDF.

The proposed MS IBI method aims to minimize the state dependence of the derived potentials
by adding additional constraints to the optimization process such that the derived
potentials fall within the region of phase space where potentials are representative of the
``true'' potential.
This approach relies on two key assumptions:
(1) different thermodynamic states have different regions of the potential phase space that
adequately reproduce their respective target RDFs, and
(2) that the true, underlying potential lies within the common overlap between these
regions of phase space.
As the name suggests, this is accomplished by updating the derived potential to
simultaneously match target RDFs at different thermodynamic states, producing a single
potential that provides sufficient matching for all target RDFs considered. 
As shown graphically in the upper portion of Figure~\ref{fig:pot-space}, the converged
potential lies at the intersection of each of the regions representing the target RDFs, as
this is the only region where a sufficient match will be found for all states.

The implementation of MS IBI is similar to that of SS IBI, the only additional requirement
is more target data.
As in SS IBI, the initial potential is assumed to be the Boltzmann inversions of the target
RDFs, averaged over the $N$ states used,
\begin{equation}
  V_0(r) = -\frac{1}{N}\sum_s k_B T_s g_s^*(r)
  \label{eq:ms-pmf}
\end{equation}
where the subscript $s$ represents the property at state $s$.
After a trial CG simulation is run at each state using the potential from
Equation~\ref{eq:ms-pmf}, the potential is updated according to:
\begin{equation}
  V_{i+1}(r) = V_i(r) + \frac{1}{N}\sum_s \alpha_s(r) k_B T_s \ln \left[
  \frac{g^i_s(r)}{g^*_s(r)} \right]
  \label{eq:ms-ibi}
\end{equation}
While in SS IBI, $\alpha$ represents a damping factor useful for suppressing fluctuations
in the potential update, here $\alpha_s(r)$ also serves as a weighting factor, allowing
more or less emphasis to be put on each state.
For example, if fitting a potential with three states, where state 1 will ultimately be of
most interest, it may make sense to give state 1 a higher $\alpha$ value; this will be
discussed later in Section~\ref{subsec:monolayers}.
Additionally, here $\alpha_s(r)$ is defined as a linear function of the separation $r$,
with the points $\alpha(0) = \alpha_{max}$, and $\alpha(r_{cutoff})=0$, such that we ensure
the derived potential has a value of zero at the interaction cutoff, $r_{cutoff}$ (i.e.,
the point at which we assume that pair interactions are zero).
Since $\alpha$ decreases as $r$ increases, increased emphasis is placed on shorter-range
interactions compared to long-range interactions, similar to the radial dependence of the
pressure correction formula often used with IBI.\cite{reith2003deriving}
This helps to suppress the influence of
long-range structural correlations on the derived potential, as short-range interactions
may certainly give rise to long-range correlations (e.g., the formation of bulk crystals
from particles interacting through a short-ranged, truncated potential). 
For direct comparability in this work, both SS and MS IBI treat the damping factor as a
linear function of separation, with a fixed value of 0 at the potential cutoff.
Note that, although bonded interactions may be optimized in a similar manner (i.e.,
adjusting the potential to match a target distribution), in this work, we make the
assumption that bonded and nonbonded interactions are sufficiently independent such that we
use analytical bonded potentials, as has been done in previous
work.\cite{hadley2010coarse,milano2005multicentered,hadley2010structurally}

The choice of states used in the fitting procedure is naturally important to deriving an
accurate potential. 
To derive the potential most representative of the underlying one, it would not be
beneficial to choose states with RDFs that are too similar, as the overlap region would be
large, essentially providing minimal additional constraints; this situation is shown in the
lower portion of Figure~\ref{fig:pot-space}.
In such a case, there would be no advantage to the multistate fitting.
At the other end of the spectrum, there may in fact be no overlap between states, or more
specifically, no overlap for a given level of matching (i.e., no overlap without relaxing
the tolerance of an RDF similarity test).
For some systems, it may not be possible to define a single pair potential that accurately
reproduces the target structure at all states. 
This is not a problem unique to CG potentials, as it applies at all levels of modeling,
e.g., classical atomistic potentials may also lack full state-independence given that they
do not allow variation in electron density.

\subsection{\label{ssec:sim-model}Simulation Model}
In this work, simulations were performed using 3 distinct models: generic LJ fluid,
TraPPE-UA, and CG models derived via IBI.
First, simulations of monatomic LJ spheres were performed in the canonical ensemble (i.e.,
fixed number of particles $N$, volume $V$, and temperature $T$), with temperature
controlled via the Nos\'{e}-Hoover thermostat.
These monatomic LJ systems contained 1468 particles initially randomly distributed
throughout the box, and were run for $1\times 10^6$ timesteps, during which the reduced
temperature was decreased from 2.0 to the final target temperature.
The systems were further equilibrated for $1\times 10^6$ timesteps before target data was
collected over $1\times 10^5$ steps.
A timestep of $1\times 10^{-3}$ reduced time units was used. 
The interaction parameters used in all LJ simulations were $\sigma=1.0$ and
$\varepsilon=1.0$, with a potential cutoff $r_{cutoff}=3\sigma$. 
Here, no coarse-graining was applied to the target systems, as these simulations were used
simply to test the efficacy of the potential derivation under the ideal circumstances where
the true potential is known.

The second model used relies on the TraPPE-UA forcefield for
simulation.\cite{martin1998transferable}
Here, alkanes were simulated in the canonical ensemble, with temperature controlled via the
Nos\'{e}-Hoover thermostat.
Bulk fluid systems for both propane (1024 molecules) and \textit{n}-dodecane (400
molecules) were simulated at 3 different states, as listed in Section~\ref{sec:results},
and used to generate target RDF data.
Although not an all-atom model (as hydrogens are not explicitly modeled), the TraPPE
forcefield was chosen for computational convenience, since, in principle, the target data
can come from any source.
In all cases, a timestep of 1 fs was used.
After an initial equilibration period of 5 ns, data was collected over a 10 ns production
run.
In addition to the bulk fluid \textit{n}-dodecane simulations, UA simulations were
performed of \textit{n}-dodecane gel and fluid monolayers, composed of 100
\textit{n}-dodecane chains each.
These were performed in the same manner as the bulk simulations at 298 K, but with the
first bead of each chain held stationary such that a 2D hexagonally arranged periodic array
with density 4.10 chains per ns$^2$ (gel) and 3.79 chains per nm$^2$ (fluid) was achieved;
these were chosen to match state points commonly used in alkylsilane monolayer simuations
and experiments.\cite{rivera2012examining}

The third model used is a CG representation of alkanes.
In all cases a 3-to-1 CG model (i.e., each CG bead represents 3 UA carbon groups) was used
to simulate bulk fluid and monolayer systems of alkanes.
Pair potentials were derived using the SS and MS IBI methods, using the results of the UA
simulations as target data, as discussed in detail in Section~\ref{sec:results}.
The bond stretching and angle bending potentials used in the study of dodecane were derived
by a Boltzmann inversion of the bonded distributions sampled in the atomistic trajectory
mapped to the CG level.\cite{milano2005multicentered}
Specifically, from a normalized bond length distribution $p(r)$, the bond stretching
potential is written as
\begin{equation}
  V_{bond}(r) = -k_B T \ln p(r)
  \label{eq:vbond}
\end{equation}
which, assuming a Gaussian bond length distribution, results in a harmonic potential about
the most probable bond length, $r_{eq}$; note an identical formalism was used for angles,
where $\theta$ is substituted for $r$, and the normalization includes a factor of
$\sin^{-1}\theta$.
Since minimal state dependence was found between systems, a single set of bonded parameters
was used in all simulations, with $k/k_B = 15.60$ K/\AA{} and $r_{eq}=3.56$ \AA{} for bonds
and $k/k_B=0.17$ K/deg$^2$ and $\theta_{eq}=174.53$ \AA{} for angles.
Bond histograms and additional details are included in the Supplemental
Material.\footnote{\label{si}See supplemental material at [URL will be inserted by AIP] for
derivation of $n$-dodecane pair and bonded potentials and associated RDFs}

In all cases, the GPU-enabled
HOOMD-Blue\cite{anderson2008general}$^,$\footnote{\lowercase{http://codeblue.umich.edu/hoomd-blue}}
simulation engine was used to perform the simulations.
The high performance of the GPU allows for rapid derivation of potentials.
A standard potential optimization using MS IBI required approximately 50 iterations to be
well-converged.
For the pure LJ systems with 1468 particles, this convergence took less than one hour using
three NVidia GTX580 GPUs concurrently.
The following convergence criteria was used to measure how well a trial RDF matched with
its target, where $dr$ is the size of an RDF bin:
\begin{equation}
  f_{fit} = 1 - \frac{\int_0^{r_{cut}} dr\, \lvert g^i(r) - g^*(r)\rvert}{
    \int_0^{r_{cut}} dr\, \lvert g^i(r) + g^*(r) \rvert}
  \label{eq:f_fit}
\end{equation}
An $f_{fit}$ value of unity represents a perfect match between the trial and target RDFs.
Additionally, in all figures, the following two-point central moving average smoothing
function was applied to the derived potential to reduce the noise:
\begin{equation}
  V^ \prime _n(r) = \frac{1}{3}\left[ V_{n-1}(r) + V_n(r) + V_{n+1}(r) \right]
  \label{eq:smooth}
\end{equation}
where $V_n(r)$ is the $n^{th}$ element of the numerical potential, and the prime denotes
the smoothed value.
The applicaiton of the smoothing function was not found to significantly influence the
behavior or degree of matching.

\section{\label{sec:results}Results}
\subsection{\label{subsec:mlj}Monatomic Lennard-Jones fluid}
To test the efficacy of the MS IBI method, potentials were derived using RDFs from
monatomic LJ spheres as target data, and the results compared to single state potential
derivation (i.e., SS IBI).
Target data was acquired from the following states: stata A, reduced density
$\rho^*=N\sigma^3/V=0.85$, reduced temperature $T^*=k_BT/\varepsilon=0.5$;
state B, $\rho^*=0.67$, $T^*=1.5$; and state C, $\rho^*=0.18$, $T^*=2.0$.
No coarse-graining was performed since the goal was to test whether the MS IBI method could
recover a known potential.
In contrast to mapping an atomistic system to the CG level, no information about the system
is lost ensuring that a single potential is applicable to all states and that this
potential is known.

\begin{figure}[h]
  \centering
  \includegraphics[width=8.5cm,keepaspectratio]{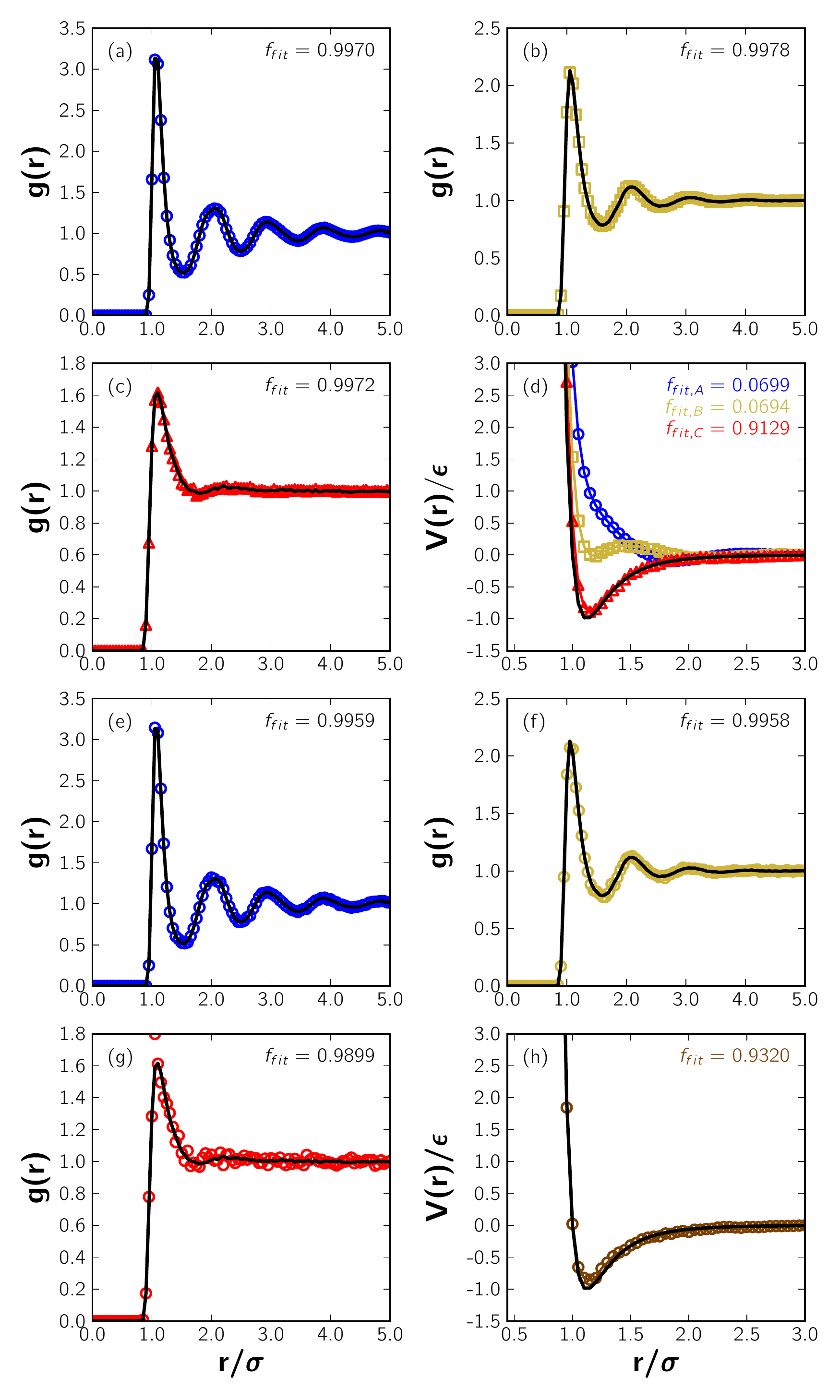}
  \caption{RDFs and potentials derived for the LJ system. (a-d) SS IBI results. (e-h) MS
  IBI results. The $\alpha$ value used for the MS IBI optimizations was 0.7 for each state.
  $f_{fit}$ for the potentials was calculated in the rancge $\sigma \leq r \leq r_{cutoff}$.
The solid black line represents the target RDF (a-c, e-g) or the known potential (d, h).
The symbols represent the derived potential (d, h) or the RDFs calculated from simulations
using the derived potential (a-c, e-g).}
  \label{fig:mlj}
\end{figure}

While the RDFs match well, as illustrated in Figure~\ref{fig:mlj}, the potentials derived
via SS IBI demonstrate significant state dependence, as shown in Figure~\ref{fig:mlj}d.
For the more dense states A and B (Figures~\ref{fig:mlj}a,b), the SS IBI method was not
able to converge to the true potential to the extent that in the most dense system (state
A), the converged potential is almost purely repulsive.
This result is due to the elevated density of this state, where the structure can be
reproduced with a purely repulsive potential.\cite{weeks1971role}
In this case, even though the RDF matches the target well, the overall behavior of the
system would be dramatically altered as compared to the target.
A similar situation arises in state B where only a weak attractive potential is required to
match the target structure.
In state C, however, the low density causes attractive forces to become important, and as
such, the attractive portion of the LJ potential is needed to fully reproduce the target
data.
Thus, the true LJ potential is recovered only for state C.
The application of SS IBI to the monatomic LJ system illustrates two points:
(1) that potentials derived via SS IBI are state-dependent, and
(2) these potentials are not unique, in that both the LJ potential and the vastly differing
derived potential produce matching RDFs.

MS IBI aims to address the aforementioned issues by forcing the potential to sample
portions of potential phase space that satisfy all of the constraints, i.e., find a single
potential that matches the target structure at multiple states.
The results of applying MS IBI to the monatomic LJ fluid are shown in
Figure~\ref{fig:mlj}e-h.
The inclusion of target data from multiple states results in closely matching RDFs and a
derived potential that accurately reproduces the true LJ potential, as shown in
Figure~\ref{fig:mlj}h.
Although this example is simple, as no coarse-graining was performed, it illustrates the
ability of MS IBI to recover a known potential and reduce the state-dependence of the derived
potential.

\subsection{\label{subsec:propane}Propane}
\begin{figure}[h]
  \centering
  \includegraphics[width=8.5cm,keepaspectratio]{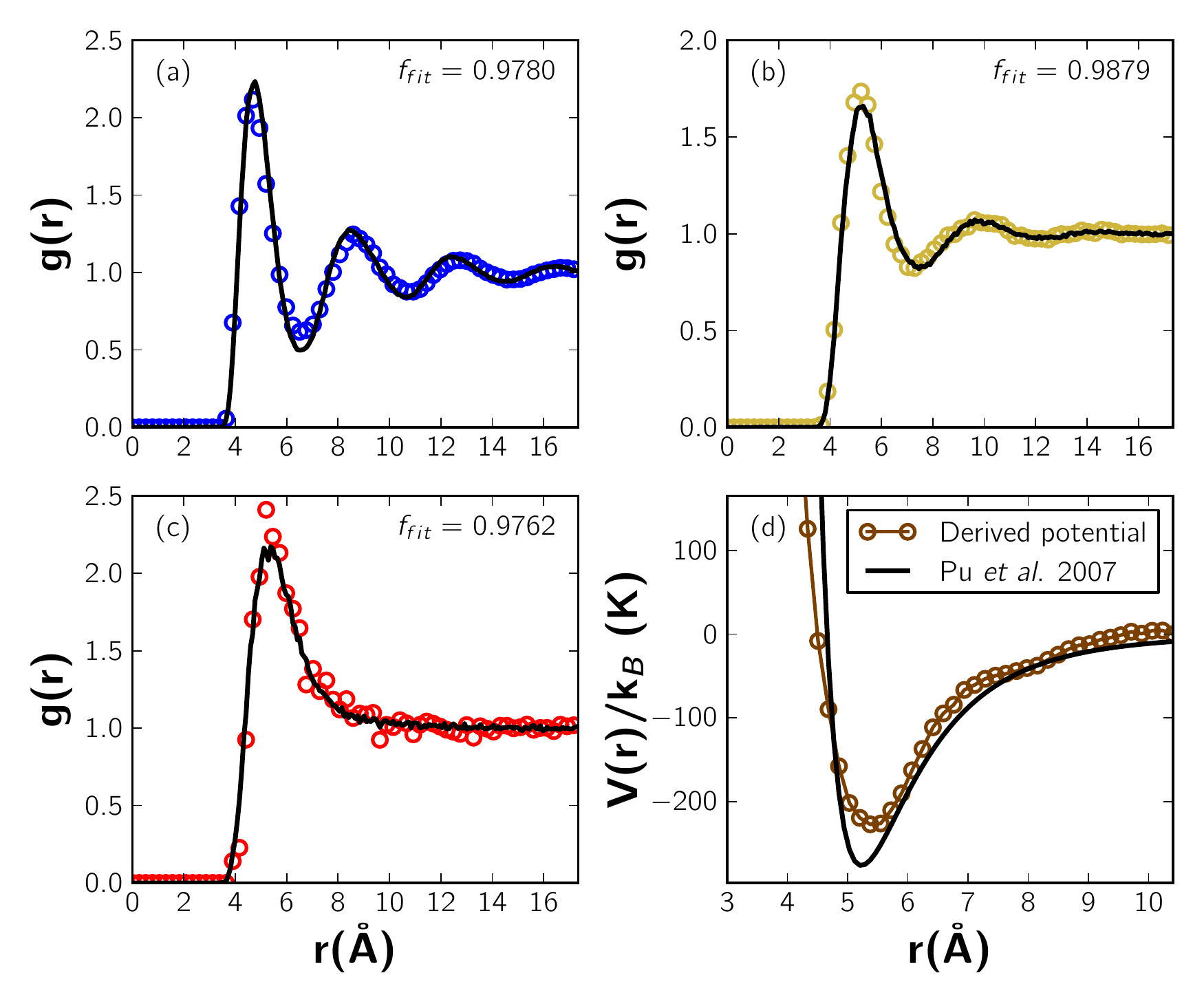}
  \caption{RDFs (a-c) and potential (d) derived for propane using MS IBI. a, b, and c
  correspond to states A, B, and C in the text, respectively. The $\alpha$ values used were
0.5, 0.7, and 0.5 for states A, B, and C, respectively.}
  \label{fig:propane}
\end{figure}

To further test the MS IBI algorithm, potential optimizations were performed on propane.
The chosen 3-to-1 mapping results in a single-site model that can be directly compared to
known single site 12-6 LJ models from the literature.\cite{pu2007molecular}
Note, the 12-6 LJ potential should not be considered to the the ``true'' potential, but
rather a good approximation.
Target data was acquired from UA simulations at the following states: state A, 298 K, 0.818
g/mL, $\alpha_A(0) = 0.5$; state B, 298 K, 0.439 g/mL, $\alpha_B(0) = 0.7$, and state C,
298 K, 0.014 g/mL, $\alpha_C(0) = 0.5$. 
The resulting RDFs and (single) pair potential are presented in Figure~\ref{fig:propane}.
At each state, $f_{fit}$ indicates excellent agreement between the target RDFs and those
calculated from simulations using the derived potential.
Moreover, we find that the derived potential agrees well with a single-site 12-6 LJ model
using parameters mapped to the critical point of propane,\cite{pu2007molecular} providing
confidence in the MS IBI method.
While the match between the two potentials is good, the derived potential does show two
small bumps at $\sim$7 \AA{} and another at $\sim$9.5 \AA{}, which are likely related to
orientational effects; that is, likely related to treating the three UA carbon groups as a
single, spherically symmetric interaction site.

\begin{figure}[h]
  \centering
  \includegraphics[width=6.5cm,keepaspectratio]{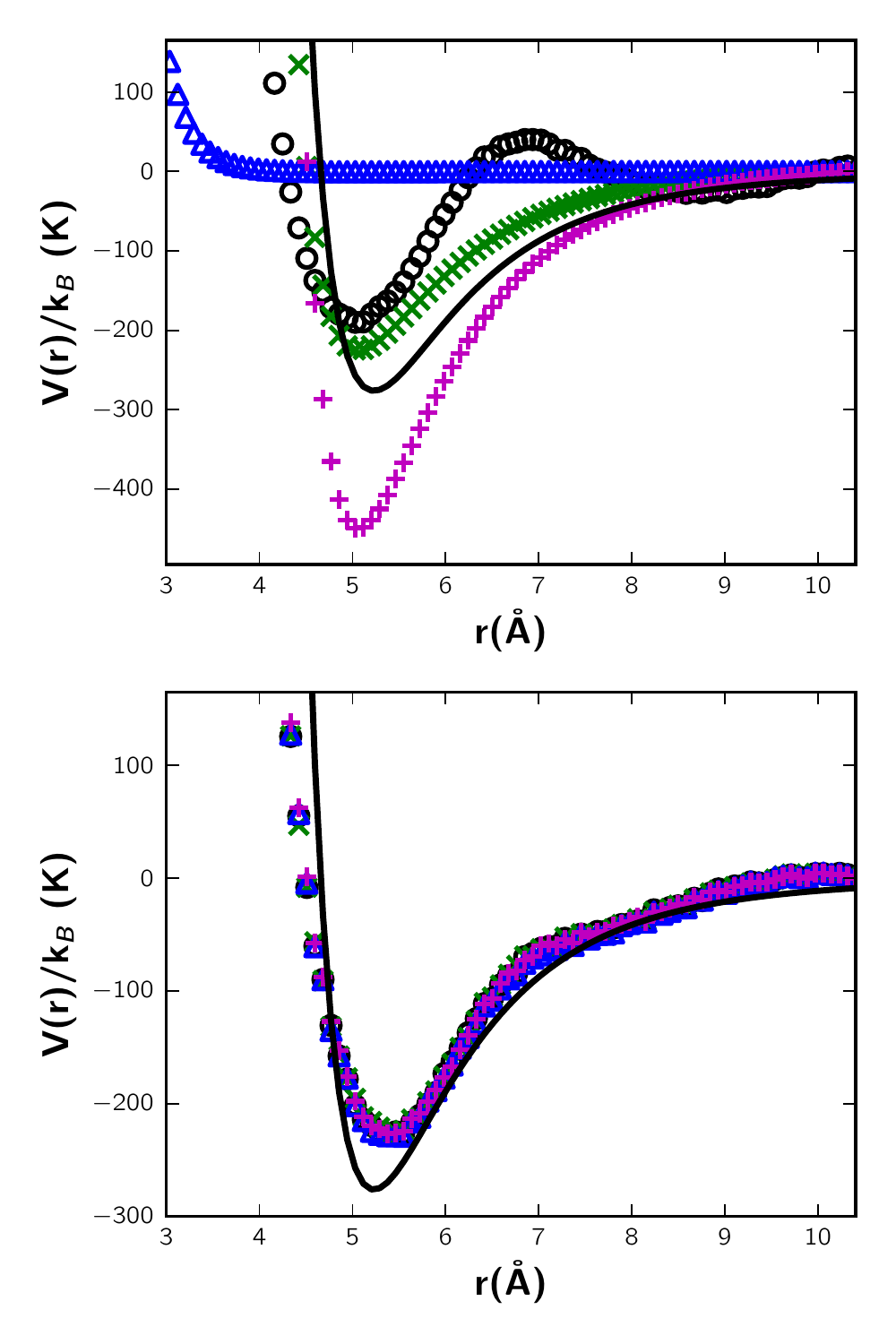}
  \caption{Different initial guesses (top) and the resulting derived potentials (bottom)
  for propane optimizations. Blue triangles: LJ with $\varepsilon=\varepsilon_2$ and
  $\sigma=\sigma_2$; black circles: $V_0(r)$ from Eq.~\ref{eq:ms-pmf} (same as shown in
  Figure~\ref{fig:propane}); green `x': LJ with $\varepsilon=\varepsilon_1$ and
  $\sigma=\sigma_1$; solid black line: 1-site propane model\cite{pu2007molecular} (not used
  as initial guess, shown for reference); magenta `$+$': LJ with $\varepsilon=\varepsilon_3$ and
  $\sigma=\sigma_3$. Symbols in top plot correspond to the same symbols in the bottom plot.
  Note that all potentials converge to very similar values.}
  \label{fig:init-pots}
\end{figure}

To illustrate the consistency of the potentials derived via MS IBI (i.e., that the final
potential is insensitive to the initial guess), optimizations were performed using a number
of different initial potentials.
In addition to the PMF-like quantity of Equation~\ref{eq:ms-pmf}, three additional initial
guesses were used, each a 12-6 LJ potential with vastly differing parameters:
(1) $\varepsilon_1 = 0.46$ kcal/mol, $\sigma_1 = 4.51$ \AA{}; 
(2) $\varepsilon_2 = 0.001$ kcal/mol, $\sigma_2 = \sigma_1$;
(3) $\varepsilon_3 = 2\varepsilon_1$, $\sigma_3 = \sigma_1$; 
The final derived potentials are, in each case, very similar to each other and to the
derived potential in Figre~\ref{fig:propane}d, as shown in Figure~\ref{fig:init-pots}.
Particularly, the $f_{fit}$ values between each potential and the derived potential shown
in Figure~\ref{fig:propane} are 0.986, 0.980, and 0.986, respectively.

\subsection{\label{subsec:dodecane}\textit{n}-dodecane}
\begin{figure}[h]
  \centering
  \includegraphics[width=5cm,keepaspectratio]{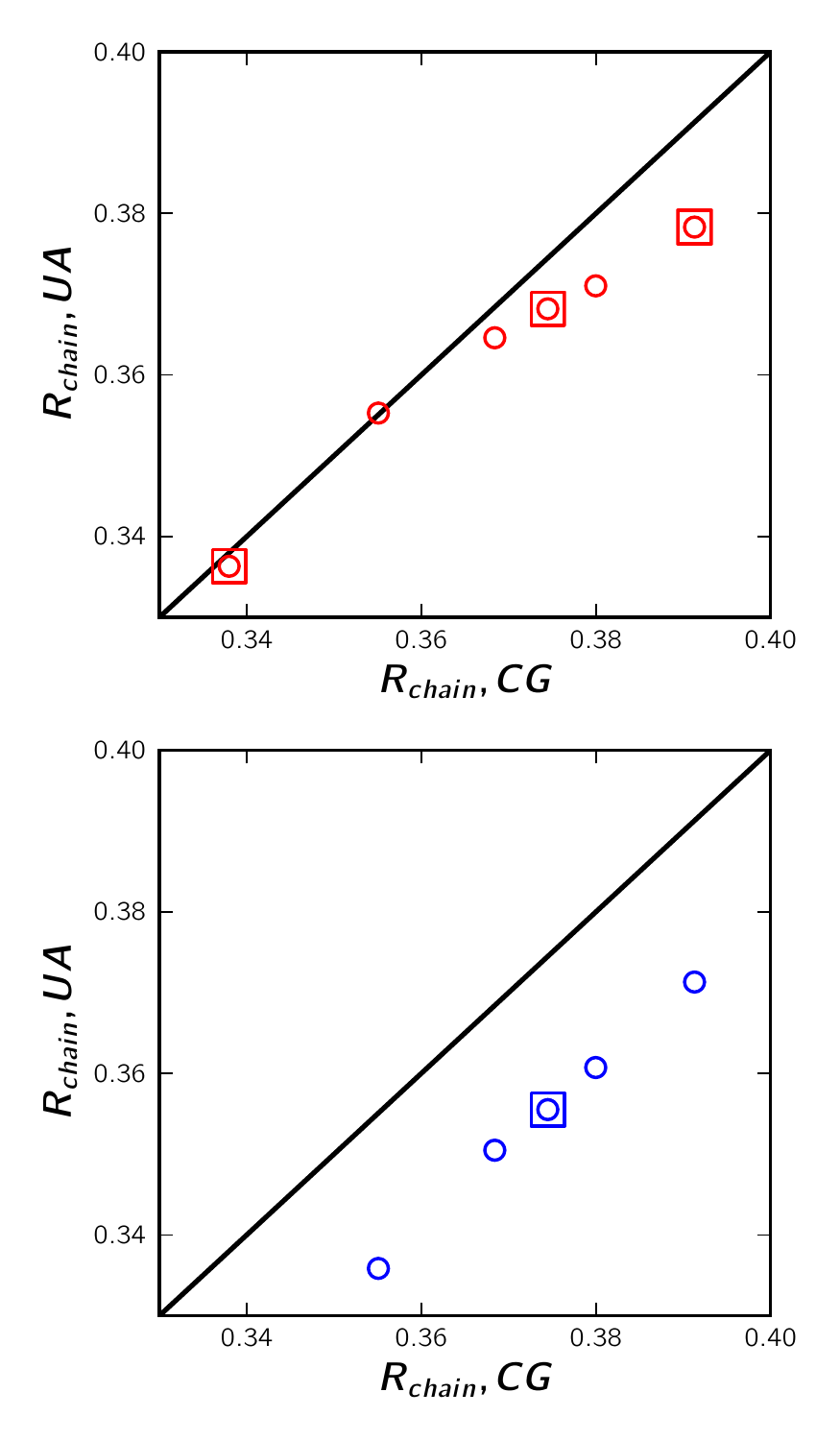}
  \caption{Comparison of a structural metric between the CG and UA models of $n$-dodecane.
  The CG potentials were derived from MS IBI (top) and SS IBI (bottom). A value lying on
the solid line represents a perfect match between the CG and UA models. Squares represent
data points from simulations at state points where the potential was derived; circles are
data points from other states used for testing the state dependence. The states used in the
multi-state fitting are states A, B, and C as described above with $\alpha_A{0} = 0.5$,
$\alpha_B(0) = 0.7$, and $\alpha_C(0) = 0.5$. State B was used for the single state fitting
with $\alpha = 0.7$.}
  \label{fig:r}
\end{figure}

To look at a more complex system and test the state-independence, we look at
\textit{n}-dodecane in the bulk and use it to examine systems containing monolayers.
Intermolecular pair potentials were derived for the beads of a CG model of
\textit{n}-dodecane, again using a 3-to-1 mapping.
The resulting 4-site model contains two middle beads and two terminal beads, where middle
and terminal beads were treated as unique entities, resulting in the need to derive three
pair potentials; harmonic bonds and angles were used, as detailed in the methods section.
The target data was collected from UA simulations of \textit{n}-dodecane at the following
states: state A, 298 K, 1.04 g/mL; state B, 298 K, 0.74 g/mL; and state C, 370 K, 0.55
g/mL; the damping values used were $\alpha_A(0) = 0.5$, $\alpha_B(0) = 0.7$, and
$\alpha_C(0) = 0.5$.
Note that state B corresponds to the experimental density at standard ambient temperature
and pressure, and, as such, is given higher weight than the other states in this example.
Close agreement with the target RDFs is found, with an $f_{fit}$ value greater than 0.98
for each of the nine RDFs calculated (not shown, see Supplemental
Material).
To further assess the quality of the potentials derived via MS IBI, the average squared
radius of gyration normalized by the average end-to-end distance, denoted by $R_{chain}$,
was calculated, providing a measure of the chain conformations at different thermodynamic
states.
Using potentials derived with MS IBI, good agreement is seen between the UA target data and
the CG model of the ratio $R_{chain}$, as shown in Figure~\ref{fig:r}; in this plotting
scheme an ideal match corresponds to a data point situated on the line $y=x$.
While deviations become more apparent as $R_{chain}$ increases, the potentials derived from
only state B (i.e., standard temperature and pressure) via SS IBI show larger, systematic
deviations of $R_{chain}$ over the entire range of state points considered.
As such, it appears the potential from MS IBI more accurately models the conformations of
dodecane over a range of state points.
Note, in both cases, additional simulations were performed at state points not used in the
fitting (state points used in the fitting are highlighted with open squares in
Figure~\ref{fig:r}), to also test the transferability of the derived potential.
The improved match of chain conformations was further tested by examining systems
containing \textit{n}-dodecane monolayers. 

\subsection{\label{subsec:monolayers}\textit{n}-dodecane Monolayer}
As mentioned in Section~\ref{subsec:ms-ibi}, here the damping coefficient, $\alpha_s(r)$,
is a function of both separation, $r$, and state, $s$. 
Recall that the dependence on separation is chosen such that the derived potential has a
value of zero at the potential cutoff as well as to reduce the influence of intermediate
and long-range structural correlations on the derived potentials.
Adjusting the $\alpha_s(0)$ value given to each state effectively alters the weight given
to each state in the fitting, i.e., more or less emphasis can be placed on a given state.
While adjusting the relative weight given to each state may have only a small effect on the
derived RDFs, it may alter the potential, which ultimately may vary other system
properties, allowing potentials to be tuned to capture specific behaviors.
To demonstrate this, as well as to further test the transferability of the derived
potentials, alkane monolayers were simulated with the 3-to-1 CG model, with potentials optimized
in the bulk states discussed above, using various values of $\alpha_s(0)$ for each of the
three states.
The average tilt angle, $\theta$, with respect to the surface and the nematic order
parameter, $S_2$, of the chains were calculated\cite{keys2011characterizing} and compared
with those values calculated from the corresponding UA simulations.
Note that the UA monolayer simulations were not used as target data in the potential
derivation, used only to validate the properties predicted by the derived CG potential.

Unique sets of CG potentials were derived over a range of $\alpha_s(0)$ values, as
summarized in Table~\ref{tab:monolayers}.
Here, the states A, B, and C are the same states previously used as target data to derive a
CG potential for bulk systems of $n$-dodecane above.
The results indicate that both the average chain tilt angle and the nematic order parameter
are functions of the relative $\alpha_s(0)$ weights for the fluid state monolayer.
For the gel phase monolayer, the nematic order parameter is less dependent on the
$\alpha_s(0)$ values, while the chain tilt angle is significantly dependent.

Initially, potentials were optimized with equal weights assigned to each state.
As shown in Table~\ref{tab:monolayers}, this $\alpha$ set yields a potential that
significantly overpredicts the fluid phase order parameter, while it underpredicts the gel
phase chain tilt.
Since the monolayers are inherently somewhat ordered, it would be expected that increasing
the relative weight given to the most dense state, state A, would yield a potential that
better captures the sytem behavior.
By systematically reducing the weight given to the less dense states, first state C, then
state B, a potential that very closely reproduces the monolayer behavior is obtained for
$\alpha$ values of 0.7, 0.1, and 0.1 for state A, B, and C, respectively.
Given the small weights assigned to states B and C in this case, it may be expected that
this potential would give results similar to the potential derived via SS IBI at state A.
However, it can be seen in Table~\ref{tab:monolayers} that this clearly is not the case;
potentials derived from SS IBI at state A show large deviations, underpredicting both the
average tilt angle and nematic order parameter in the gel phase monolayer, in stark
contrast to the near perfect behavior predicted by MS IBI.
This result is a direct consequence of using the MS IBI method; even though low weights are
given to the other states, the derived potentials will only be considered converged if all
states demonstrate good agreement.
Again, we note that when deriving the potentials, the UA monolayer was not used as target
data (i.e., the structure matching was performed in the same manner as described in
Section~\ref{subsec:dodecane}, except with varying values of $\alpha_s(0)$ for each state).
The close match that is observed is a result of the success of the MS IBI method in
deriving a more generally applicable, transferable, set of potentials.
\begin{table*}
  \centering
  \caption{Average chain tilt with respect to surface normal, $\theta$, and nematic order
  parameter, $S_2$, for the monolayers in the fluid state (subscript F) and in the gel
state (subscript G). The states A, B, and C are the same ones used in
Section~\ref{subsec:dodecane}. Values are given as ensemble averages $\pm$ standard
deviation.}
  \begin{tabular}{@{} l l l l l @{}}
    \toprule
    $\alpha_A$ / $\alpha_B$ / $\alpha_C\,$ & $\theta_F$ & $S_{2,F}$ & $\theta_G$ & $S_{2,G}$ \\
    \hline
    0.7/0.7/0.7 & $19\pm6.3^{\circ}$ & $0.97\pm0.013$ & $18\pm6.5^{\circ}$ & $0.98\pm0.017$ \\
    0.7/0.7/0.6 & $20\pm6.1^{\circ}$ & $0.97\pm0.012$ & $20\pm3.7^{\circ}$ & $0.992\pm0.0018$ \\
    0.7/0.7/0.5 & $18\pm6.9^{\circ}$ & $0.96\pm0.020$ & $20\pm3.8^{\circ}$ & $0.991\pm0.0022$ \\
    0.7/0.7/0.4 & $15\pm7.0^{\circ}$ & $0.95\pm0.018$ & $20\pm4.0^{\circ}$ & $0.990\pm0.0027$ \\
    0.7/0.7/0.3 & $13\pm6.8^{\circ}$ & $0.94\pm0.019$ & $19\pm4.2^{\circ}$ & $0.990\pm0.0036$ \\
    0.7/0.7/0.2 & $12\pm6.3^{\circ}$ & $0.93\pm0.014$ & $12\pm6.4^{\circ}$ & $0.96\pm0.014$ \\
    0.7/0.7/0.1 & $10\pm5.6^{\circ}$ & $0.95\pm0.011$ & $8 \pm5.3^{\circ}$ & $0.958\pm0.008$ \\
    0.7/0.4/0.1 & $18\pm9.0^{\circ}$ & $0.83\pm0.021$ & $29\pm3.9^{\circ}$ & $0.988\pm0.0026$ \\
    0.7/0.1/0.1 & $20\pm10^{\circ}$  & $0.81\pm0.023$ & $32.2\pm3.6^{\circ}$ & $0.989\pm0.002$ \\
    \hline
    SS IBI, state A & $17\pm8.8^{\circ}$ & $0.85\pm0.017$ & $15\pm7.4^{\circ}$ & $0.89\pm0.014$\\
    SS IBI, state B & $23\pm4.9^{\circ}$ & $0.983\pm0.005$ & $21.3\pm3.3^{\circ}$ & $0.993\pm0.001$\\
    SS IBI, state C & $23\pm4.1^{\circ}$ & $0.988\pm0.002$ & $20\pm4.6^{\circ}$ & $0.987\pm0.002$\\
    \hline
    United Atom & $20\pm15^{\circ}$ & $0.820\pm0.028$ & $32.4\pm10.1^{\circ}$ & $0.965\pm0.046$\\
    \toprule
  \end{tabular}
  \label{tab:monolayers}
\end{table*}

\section{\label{sec:conclusion}Conclusion}
A multistate extension of the popular IBI method has been proposed.
In the proposed MS IBI method, multiple thermodynamic states are used in the derivation of a
single, generally applicable potential.
For systems with a known potential, it was shown that the MS IBI method was capable of accurately
recovering the true, underlying potential, while the SS IBI method was unable to consistently
derive a gnerally applicable potential.
Through the coarse-graining of propane, it was shown that MS IBI was able to recover a potential
very similar to a previously published single-site model with good reproducibility.
Furthermore, potentials derived via MS IBI were shown to better reproduce structural conformations
of $n$-dodecane than potentials derived via SS IBI.
It was also demonstrated that adjusting the relative weights given to each target in the
optimizations can be used to tune system properties beyond the RDF; in this case, tuning the
weights enabled potentials to be derived that provided near perfect agreement between CG and
atomistic models when considering the nematic order parameter and tilt angle of an $n$-dodecane
monolayer.
While pressure, and thermodynamics in general, were not investigated in this work, the standard
pressure correction scheme of SS IBI\cite{reith2003deriving} could be trivially applied to MS IBI
by calculating the average pressure deviations between all states, and using this quantity in the
pressure correction term.
As such, the MS IBI stands as an improvement of the typical IBI method, producing more generally
applicable potentials that can be tuned to match target properties from experiment or
finer-grained simulations.

This improved methodology should be very useful for a host of molecular systems, including,
for example, lipid systems, where not only do systems demonstrate structural heterogeneity
within a given state point (i.e., different molecular structures in a single system), but
properties such as tilt angle, nematic order, area per lipid, etc., need to be tuned in
order to match atomistic simulations and experiment.
\cite{heller1993molecular,egberts1994molecular,saiz2002computer,scott2002modeling,venturoli2006mesoscopic}
Given that the MS IBI approach is also capable of deriving potentials which demonstrate
increased levels of transferability than SS IBI, potentials can be derived for complex
systems with many unique interactions by examining the individual components separately,
reducing the number of simultaneous optimizations that need to be performed.
Furthermore, this work presents a method to develop potentials that enable the examination
of phase transitions; in many prior works utilizing SS IBI, different potentials are needed
to appropriately model different states, making it difficult to accurately examine the
transition between those states.
\cite{marrink2007martini,hadley2010coarse,vettorel2006coarse,d2012coarse}
Additionally, given that multi-GPU machines and GPU enabled simulation
packages\cite{anderson2008general,hess2008gromacs,plimpton1995fast,brown2011implementing,brown2012implementing}
are becoming more common, the potential derivation process can be carried
out with relatively little computational effort, even if a large number of targets are
needed, or a large number of iterations must be undertaken to find appropriate weighting
functions.

\acknowledgments{
  The authors acknowledge support from grant number R01 AR057886-01 from the National
Institute of Arthritis and Musculoskeletal and Skin Diseases.}

\bibliography{ms-ibi}

\begin{thebibliography}{72}%
\makeatletter
\providecommand \@ifxundefined [1]{%
 \@ifx{#1\undefined}
}%
\providecommand \@ifnum [1]{%
 \ifnum #1\expandafter \@firstoftwo
 \else \expandafter \@secondoftwo
 \fi
}%
\providecommand \@ifx [1]{%
 \ifx #1\expandafter \@firstoftwo
 \else \expandafter \@secondoftwo
 \fi
}%
\providecommand \natexlab [1]{#1}%
\providecommand \enquote  [1]{``#1''}%
\providecommand \bibnamefont  [1]{#1}%
\providecommand \bibfnamefont [1]{#1}%
\providecommand \citenamefont [1]{#1}%
\providecommand \href@noop [0]{\@secondoftwo}%
\providecommand \href [0]{\begingroup \@sanitize@url \@href}%
\providecommand \@href[1]{\@@startlink{#1}\@@href}%
\providecommand \@@href[1]{\endgroup#1\@@endlink}%
\providecommand \@sanitize@url [0]{\catcode `\\12\catcode `\$12\catcode
  `\&12\catcode `\#12\catcode `\^12\catcode `\_12\catcode `\%12\relax}%
\providecommand \@@startlink[1]{}%
\providecommand \@@endlink[0]{}%
\providecommand \url  [0]{\begingroup\@sanitize@url \@url }%
\providecommand \@url [1]{\endgroup\@href {#1}{\urlprefix }}%
\providecommand \urlprefix  [0]{URL }%
\providecommand \Eprint [0]{\href }%
\providecommand \doibase [0]{http://dx.doi.org/}%
\providecommand \selectlanguage [0]{\@gobble}%
\providecommand \bibinfo  [0]{\@secondoftwo}%
\providecommand \bibfield  [0]{\@secondoftwo}%
\providecommand \translation [1]{[#1]}%
\providecommand \BibitemOpen [0]{}%
\providecommand \bibitemStop [0]{}%
\providecommand \bibitemNoStop [0]{.\EOS\space}%
\providecommand \EOS [0]{\spacefactor3000\relax}%
\providecommand \BibitemShut  [1]{\csname bibitem#1\endcsname}%
\let\auto@bib@innerbib\@empty
\bibitem [{\citenamefont {Israelachvili}\ \emph {et~al.}(1976)\citenamefont
  {Israelachvili}, \citenamefont {Mitchell},\ and\ \citenamefont
  {Ninham}}]{israelachvili1976theory}%
  \BibitemOpen
  \bibfield  {author} {\bibinfo {author} {\bibfnamefont {J.~N.}\ \bibnamefont
  {Israelachvili}}, \bibinfo {author} {\bibfnamefont {D.~J.}\ \bibnamefont
  {Mitchell}}, \ and\ \bibinfo {author} {\bibfnamefont {B.~W.}\ \bibnamefont
  {Ninham}},\ }\href@noop {} {\bibfield  {journal} {\bibinfo  {journal} {J.
  Chem. Soc., Faraday Trans. 2}\ }\textbf {\bibinfo {volume} {72}},\ \bibinfo
  {pages} {1525} (\bibinfo {year} {1976})}\BibitemShut {NoStop}%
\bibitem [{\citenamefont {Israelachvili}\ \emph {et~al.}(1977)\citenamefont
  {Israelachvili}, \citenamefont {Mitchell},\ and\ \citenamefont
  {Ninham}}]{israelachvili1977theory}%
  \BibitemOpen
  \bibfield  {author} {\bibinfo {author} {\bibfnamefont {J.~N.}\ \bibnamefont
  {Israelachvili}}, \bibinfo {author} {\bibfnamefont {D.~J.}\ \bibnamefont
  {Mitchell}}, \ and\ \bibinfo {author} {\bibfnamefont {B.~W.}\ \bibnamefont
  {Ninham}},\ }\href@noop {} {\bibfield  {journal} {\bibinfo  {journal}
  {Biochimica et Biophysica Acta (BBA)-Biomembranes}\ }\textbf {\bibinfo
  {volume} {470}},\ \bibinfo {pages} {185} (\bibinfo {year}
  {1977})}\BibitemShut {NoStop}%
\bibitem [{\citenamefont {Mu{\v{s}}evi{\v{c}}}\ and\ \citenamefont
  {{\v{S}}karabot}(2008)}]{muvsevivc2008self}%
  \BibitemOpen
  \bibfield  {author} {\bibinfo {author} {\bibfnamefont {I.}~\bibnamefont
  {Mu{\v{s}}evi{\v{c}}}}\ and\ \bibinfo {author} {\bibfnamefont
  {M.}~\bibnamefont {{\v{S}}karabot}},\ }\href@noop {} {\bibfield  {journal}
  {\bibinfo  {journal} {Soft Matter}\ }\textbf {\bibinfo {volume} {4}},\
  \bibinfo {pages} {195} (\bibinfo {year} {2008})}\BibitemShut {NoStop}%
\bibitem [{\citenamefont {Bishop}\ \emph {et~al.}(2009)\citenamefont {Bishop},
  \citenamefont {Wilmer}, \citenamefont {Soh},\ and\ \citenamefont
  {Grzybowski}}]{bishop2009nanoscale}%
  \BibitemOpen
  \bibfield  {author} {\bibinfo {author} {\bibfnamefont {K.~J.}\ \bibnamefont
  {Bishop}}, \bibinfo {author} {\bibfnamefont {C.~E.}\ \bibnamefont {Wilmer}},
  \bibinfo {author} {\bibfnamefont {S.}~\bibnamefont {Soh}}, \ and\ \bibinfo
  {author} {\bibfnamefont {B.~A.}\ \bibnamefont {Grzybowski}},\ }\href@noop {}
  {\bibfield  {journal} {\bibinfo  {journal} {small}\ }\textbf {\bibinfo
  {volume} {5}},\ \bibinfo {pages} {1600} (\bibinfo {year} {2009})}\BibitemShut
  {NoStop}%
\bibitem [{\citenamefont {Iacovella}\ \emph {et~al.}(2011)\citenamefont
  {Iacovella}, \citenamefont {Keys},\ and\ \citenamefont
  {Glotzer}}]{iacovella2011self}%
  \BibitemOpen
  \bibfield  {author} {\bibinfo {author} {\bibfnamefont {C.~R.}\ \bibnamefont
  {Iacovella}}, \bibinfo {author} {\bibfnamefont {A.~S.}\ \bibnamefont {Keys}},
  \ and\ \bibinfo {author} {\bibfnamefont {S.~C.}\ \bibnamefont {Glotzer}},\
  }\href@noop {} {\bibfield  {journal} {\bibinfo  {journal} {Proc. of the Nat.
  Ac. of Sci.}\ }\textbf {\bibinfo {volume} {108}},\ \bibinfo {pages} {20935}
  (\bibinfo {year} {2011})}\BibitemShut {NoStop}%
\bibitem [{\citenamefont {Ziherl}\ and\ \citenamefont
  {Kamien}(2001)}]{ziherl2001maximizing}%
  \BibitemOpen
  \bibfield  {author} {\bibinfo {author} {\bibfnamefont {P.}~\bibnamefont
  {Ziherl}}\ and\ \bibinfo {author} {\bibfnamefont {R.~D.}\ \bibnamefont
  {Kamien}},\ }\href@noop {} {\bibfield  {journal} {\bibinfo  {journal} {J.
  Phys. Chem. B}\ }\textbf {\bibinfo {volume} {105}},\ \bibinfo {pages} {10147}
  (\bibinfo {year} {2001})}\BibitemShut {NoStop}%
\bibitem [{\citenamefont {Thomas}\ \emph {et~al.}(1987)\citenamefont {Thomas},
  \citenamefont {Kinning}, \citenamefont {Alward},\ and\ \citenamefont
  {Henkee}}]{thomas1987ordered}%
  \BibitemOpen
  \bibfield  {author} {\bibinfo {author} {\bibfnamefont {E.~L.}\ \bibnamefont
  {Thomas}}, \bibinfo {author} {\bibfnamefont {D.~J.}\ \bibnamefont {Kinning}},
  \bibinfo {author} {\bibfnamefont {D.~B.}\ \bibnamefont {Alward}}, \ and\
  \bibinfo {author} {\bibfnamefont {C.~S.}\ \bibnamefont {Henkee}},\
  }\href@noop {} {\bibfield  {journal} {\bibinfo  {journal} {Macromol.}\
  }\textbf {\bibinfo {volume} {20}},\ \bibinfo {pages} {2934} (\bibinfo {year}
  {1987})}\BibitemShut {NoStop}%
\bibitem [{\citenamefont {Phillips}\ \emph {et~al.}(2010)\citenamefont
  {Phillips}, \citenamefont {Iacovella},\ and\ \citenamefont
  {Glotzer}}]{phillips2010stability}%
  \BibitemOpen
  \bibfield  {author} {\bibinfo {author} {\bibfnamefont {C.~L.}\ \bibnamefont
  {Phillips}}, \bibinfo {author} {\bibfnamefont {C.~R.}\ \bibnamefont
  {Iacovella}}, \ and\ \bibinfo {author} {\bibfnamefont {S.~C.}\ \bibnamefont
  {Glotzer}},\ }\href@noop {} {\bibfield  {journal} {\bibinfo  {journal} {Soft
  Matter}\ }\textbf {\bibinfo {volume} {6}},\ \bibinfo {pages} {1693} (\bibinfo
  {year} {2010})}\BibitemShut {NoStop}%
\bibitem [{\citenamefont {Iacovella}\ and\ \citenamefont
  {Glotzer}(2009)}]{iacovella2009complex}%
  \BibitemOpen
  \bibfield  {author} {\bibinfo {author} {\bibfnamefont {C.~R.}\ \bibnamefont
  {Iacovella}}\ and\ \bibinfo {author} {\bibfnamefont {S.~C.}\ \bibnamefont
  {Glotzer}},\ }\href@noop {} {\bibfield  {journal} {\bibinfo  {journal} {Nano
  Lett.}\ }\textbf {\bibinfo {volume} {9}},\ \bibinfo {pages} {1206} (\bibinfo
  {year} {2009})}\BibitemShut {NoStop}%
\bibitem [{\citenamefont {Glaser}\ \emph {et~al.}(2007)\citenamefont {Glaser},
  \citenamefont {Grason}, \citenamefont {Kamien}, \citenamefont
  {Ko{\v{s}}mrlj}, \citenamefont {Santangelo},\ and\ \citenamefont
  {Ziherl}}]{glaser2007soft}%
  \BibitemOpen
  \bibfield  {author} {\bibinfo {author} {\bibfnamefont {M.~A.}\ \bibnamefont
  {Glaser}}, \bibinfo {author} {\bibfnamefont {G.~M.}\ \bibnamefont {Grason}},
  \bibinfo {author} {\bibfnamefont {R.~D.}\ \bibnamefont {Kamien}}, \bibinfo
  {author} {\bibfnamefont {A.}~\bibnamefont {Ko{\v{s}}mrlj}}, \bibinfo {author}
  {\bibfnamefont {C.~D.}\ \bibnamefont {Santangelo}}, \ and\ \bibinfo {author}
  {\bibfnamefont {P.}~\bibnamefont {Ziherl}},\ }\href@noop {} {\bibfield
  {journal} {\bibinfo  {journal} {Europhys. Lett.}\ }\textbf {\bibinfo {volume}
  {78}},\ \bibinfo {pages} {46004} (\bibinfo {year} {2007})}\BibitemShut
  {NoStop}%
\bibitem [{\citenamefont {Ziherl}\ and\ \citenamefont
  {Kamien}(2000)}]{ziherl2000soap}%
  \BibitemOpen
  \bibfield  {author} {\bibinfo {author} {\bibfnamefont {P.}~\bibnamefont
  {Ziherl}}\ and\ \bibinfo {author} {\bibfnamefont {R.~D.}\ \bibnamefont
  {Kamien}},\ }\href@noop {} {\bibfield  {journal} {\bibinfo  {journal} {Phys.
  Rev. Lett.}\ }\textbf {\bibinfo {volume} {85}},\ \bibinfo {pages} {3528}
  (\bibinfo {year} {2000})}\BibitemShut {NoStop}%
\bibitem [{\citenamefont {Larson}\ \emph {et~al.}(1985)\citenamefont {Larson},
  \citenamefont {Scriven},\ and\ \citenamefont {Davis}}]{larson1985monte}%
  \BibitemOpen
  \bibfield  {author} {\bibinfo {author} {\bibfnamefont {R.}~\bibnamefont
  {Larson}}, \bibinfo {author} {\bibfnamefont {L.}~\bibnamefont {Scriven}}, \
  and\ \bibinfo {author} {\bibfnamefont {H.}~\bibnamefont {Davis}},\
  }\href@noop {} {\bibfield  {journal} {\bibinfo  {journal} {J. Chem. Phys.}\
  }\textbf {\bibinfo {volume} {83}},\ \bibinfo {pages} {2411} (\bibinfo {year}
  {1985})}\BibitemShut {NoStop}%
\bibitem [{\citenamefont {Glotzer}\ \emph {et~al.}(2005)\citenamefont
  {Glotzer}, \citenamefont {Horsch}, \citenamefont {Iacovella}, \citenamefont
  {Zhang}, \citenamefont {Chan},\ and\ \citenamefont
  {Zhang}}]{glotzer2005self}%
  \BibitemOpen
  \bibfield  {author} {\bibinfo {author} {\bibfnamefont {S.~C.}\ \bibnamefont
  {Glotzer}}, \bibinfo {author} {\bibfnamefont {M.~A.}\ \bibnamefont {Horsch}},
  \bibinfo {author} {\bibfnamefont {C.~R.}\ \bibnamefont {Iacovella}}, \bibinfo
  {author} {\bibfnamefont {Z.}~\bibnamefont {Zhang}}, \bibinfo {author}
  {\bibfnamefont {E.~R.}\ \bibnamefont {Chan}}, \ and\ \bibinfo {author}
  {\bibfnamefont {X.}~\bibnamefont {Zhang}},\ }\href@noop {} {\bibfield
  {journal} {\bibinfo  {journal} {Current Opinion in Colloid \& Interface
  Science}\ }\textbf {\bibinfo {volume} {10}},\ \bibinfo {pages} {287}
  (\bibinfo {year} {2005})}\BibitemShut {NoStop}%
\bibitem [{\citenamefont {Gai}\ \emph {et~al.}(2012)\citenamefont {Gai},
  \citenamefont {Maerzke}, \citenamefont {Cummings},\ and\ \citenamefont
  {McCabe}}]{gai2012wang}%
  \BibitemOpen
  \bibfield  {author} {\bibinfo {author} {\bibfnamefont {L.}~\bibnamefont
  {Gai}}, \bibinfo {author} {\bibfnamefont {K.~A.}\ \bibnamefont {Maerzke}},
  \bibinfo {author} {\bibfnamefont {P.~T.}\ \bibnamefont {Cummings}}, \ and\
  \bibinfo {author} {\bibfnamefont {C.}~\bibnamefont {McCabe}},\ }\href@noop {}
  {\bibfield  {journal} {\bibinfo  {journal} {J. Chem. Phys.}\ }\textbf
  {\bibinfo {volume} {137}},\ \bibinfo {pages} {144901} (\bibinfo {year}
  {2012})}\BibitemShut {NoStop}%
\bibitem [{\citenamefont {Iacovella}\ \emph {et~al.}(2005)\citenamefont
  {Iacovella}, \citenamefont {Horsch}, \citenamefont {Zhang},\ and\
  \citenamefont {Glotzer}}]{iacovella2005phase}%
  \BibitemOpen
  \bibfield  {author} {\bibinfo {author} {\bibfnamefont {C.~R.}\ \bibnamefont
  {Iacovella}}, \bibinfo {author} {\bibfnamefont {M.~A.}\ \bibnamefont
  {Horsch}}, \bibinfo {author} {\bibfnamefont {Z.}~\bibnamefont {Zhang}}, \
  and\ \bibinfo {author} {\bibfnamefont {S.~C.}\ \bibnamefont {Glotzer}},\
  }\href@noop {} {\bibfield  {journal} {\bibinfo  {journal} {Langmuir}\
  }\textbf {\bibinfo {volume} {21}},\ \bibinfo {pages} {9488} (\bibinfo {year}
  {2005})}\BibitemShut {NoStop}%
\bibitem [{\citenamefont {Lau}\ and\ \citenamefont
  {Dill}(1989)}]{lau1989lattice}%
  \BibitemOpen
  \bibfield  {author} {\bibinfo {author} {\bibfnamefont {K.~F.}\ \bibnamefont
  {Lau}}\ and\ \bibinfo {author} {\bibfnamefont {K.~A.}\ \bibnamefont {Dill}},\
  }\href@noop {} {\bibfield  {journal} {\bibinfo  {journal} {Macromol.}\
  }\textbf {\bibinfo {volume} {22}},\ \bibinfo {pages} {3986} (\bibinfo {year}
  {1989})}\BibitemShut {NoStop}%
\bibitem [{\citenamefont {Maerzke}\ and\ \citenamefont
  {Siepmann}(2011)}]{maerzke2011transferable}%
  \BibitemOpen
  \bibfield  {author} {\bibinfo {author} {\bibfnamefont {K.~A.}\ \bibnamefont
  {Maerzke}}\ and\ \bibinfo {author} {\bibfnamefont {J.~I.}\ \bibnamefont
  {Siepmann}},\ }\href@noop {} {\bibfield  {journal} {\bibinfo  {journal} {J.
  Phys. Chem. B}\ }\textbf {\bibinfo {volume} {115}},\ \bibinfo {pages} {3452}
  (\bibinfo {year} {2011})}\BibitemShut {NoStop}%
\bibitem [{\citenamefont {Marrink}\ \emph {et~al.}(2007)\citenamefont
  {Marrink}, \citenamefont {Risselada}, \citenamefont {Yefimov}, \citenamefont
  {Tieleman},\ and\ \citenamefont {de~Vries}}]{marrink2007martini}%
  \BibitemOpen
  \bibfield  {author} {\bibinfo {author} {\bibfnamefont {S.~J.}\ \bibnamefont
  {Marrink}}, \bibinfo {author} {\bibfnamefont {H.~J.}\ \bibnamefont
  {Risselada}}, \bibinfo {author} {\bibfnamefont {S.}~\bibnamefont {Yefimov}},
  \bibinfo {author} {\bibfnamefont {D.~P.}\ \bibnamefont {Tieleman}}, \ and\
  \bibinfo {author} {\bibfnamefont {A.~H.}\ \bibnamefont {de~Vries}},\
  }\href@noop {} {\bibfield  {journal} {\bibinfo  {journal} {J. Phys. Chem. B}\
  }\textbf {\bibinfo {volume} {111}},\ \bibinfo {pages} {7812} (\bibinfo {year}
  {2007})}\BibitemShut {NoStop}%
\bibitem [{\citenamefont {Martin}\ and\ \citenamefont
  {Siepmann}(1998)}]{martin1998transferable}%
  \BibitemOpen
  \bibfield  {author} {\bibinfo {author} {\bibfnamefont {M.~G.}\ \bibnamefont
  {Martin}}\ and\ \bibinfo {author} {\bibfnamefont {J.~I.}\ \bibnamefont
  {Siepmann}},\ }\href@noop {} {\bibfield  {journal} {\bibinfo  {journal} {J.
  Phys. Chem. B}\ }\textbf {\bibinfo {volume} {102}},\ \bibinfo {pages} {2569}
  (\bibinfo {year} {1998})}\BibitemShut {NoStop}%
\bibitem [{\citenamefont {Jorgensen}\ \emph {et~al.}(1996)\citenamefont
  {Jorgensen}, \citenamefont {Maxwell},\ and\ \citenamefont
  {Tirado-Rives}}]{jorgensen1996development}%
  \BibitemOpen
  \bibfield  {author} {\bibinfo {author} {\bibfnamefont {W.~L.}\ \bibnamefont
  {Jorgensen}}, \bibinfo {author} {\bibfnamefont {D.~S.}\ \bibnamefont
  {Maxwell}}, \ and\ \bibinfo {author} {\bibfnamefont {J.}~\bibnamefont
  {Tirado-Rives}},\ }\href@noop {} {\bibfield  {journal} {\bibinfo  {journal}
  {J. Am. Chem. Soc.}\ }\textbf {\bibinfo {volume} {118}},\ \bibinfo {pages}
  {11225} (\bibinfo {year} {1996})}\BibitemShut {NoStop}%
\bibitem [{\citenamefont {Oostenbrink}\ \emph {et~al.}(2004)\citenamefont
  {Oostenbrink}, \citenamefont {Villa}, \citenamefont {Mark},\ and\
  \citenamefont {Van~Gunsteren}}]{oostenbrink2004biomolecular}%
  \BibitemOpen
  \bibfield  {author} {\bibinfo {author} {\bibfnamefont {C.}~\bibnamefont
  {Oostenbrink}}, \bibinfo {author} {\bibfnamefont {A.}~\bibnamefont {Villa}},
  \bibinfo {author} {\bibfnamefont {A.~E.}\ \bibnamefont {Mark}}, \ and\
  \bibinfo {author} {\bibfnamefont {W.~F.}\ \bibnamefont {Van~Gunsteren}},\
  }\href@noop {} {\bibfield  {journal} {\bibinfo  {journal} {J. Comp. Chem.}\
  }\textbf {\bibinfo {volume} {25}},\ \bibinfo {pages} {1656} (\bibinfo {year}
  {2004})}\BibitemShut {NoStop}%
\bibitem [{\citenamefont {Vanommeslaeghe}\ \emph {et~al.}(2010)\citenamefont
  {Vanommeslaeghe}, \citenamefont {Hatcher}, \citenamefont {Acharya},
  \citenamefont {Kundu}, \citenamefont {Zhong}, \citenamefont {Shim},
  \citenamefont {Darian}, \citenamefont {Guvench}, \citenamefont {Lopes},
  \citenamefont {Vorobyov} \emph {et~al.}}]{vanommeslaeghe2010charmm}%
  \BibitemOpen
  \bibfield  {author} {\bibinfo {author} {\bibfnamefont {K.}~\bibnamefont
  {Vanommeslaeghe}}, \bibinfo {author} {\bibfnamefont {E.}~\bibnamefont
  {Hatcher}}, \bibinfo {author} {\bibfnamefont {C.}~\bibnamefont {Acharya}},
  \bibinfo {author} {\bibfnamefont {S.}~\bibnamefont {Kundu}}, \bibinfo
  {author} {\bibfnamefont {S.}~\bibnamefont {Zhong}}, \bibinfo {author}
  {\bibfnamefont {J.}~\bibnamefont {Shim}}, \bibinfo {author} {\bibfnamefont
  {E.}~\bibnamefont {Darian}}, \bibinfo {author} {\bibfnamefont
  {O.}~\bibnamefont {Guvench}}, \bibinfo {author} {\bibfnamefont
  {P.}~\bibnamefont {Lopes}}, \bibinfo {author} {\bibfnamefont
  {I.}~\bibnamefont {Vorobyov}},  \emph {et~al.},\ }\href@noop {} {\bibfield
  {journal} {\bibinfo  {journal} {J. Comp. Chem.}\ }\textbf {\bibinfo {volume}
  {31}},\ \bibinfo {pages} {671} (\bibinfo {year} {2010})}\BibitemShut
  {NoStop}%
\bibitem [{\citenamefont {Wang}\ \emph {et~al.}(2004)\citenamefont {Wang},
  \citenamefont {Wolf}, \citenamefont {Caldwell}, \citenamefont {Kollman},\
  and\ \citenamefont {Case}}]{wang2004development}%
  \BibitemOpen
  \bibfield  {author} {\bibinfo {author} {\bibfnamefont {J.}~\bibnamefont
  {Wang}}, \bibinfo {author} {\bibfnamefont {R.~M.}\ \bibnamefont {Wolf}},
  \bibinfo {author} {\bibfnamefont {J.~W.}\ \bibnamefont {Caldwell}}, \bibinfo
  {author} {\bibfnamefont {P.~A.}\ \bibnamefont {Kollman}}, \ and\ \bibinfo
  {author} {\bibfnamefont {D.~A.}\ \bibnamefont {Case}},\ }\href@noop {}
  {\bibfield  {journal} {\bibinfo  {journal} {J. Comp. Chem.}\ }\textbf
  {\bibinfo {volume} {25}},\ \bibinfo {pages} {1157} (\bibinfo {year}
  {2004})}\BibitemShut {NoStop}%
\bibitem [{\citenamefont {McCabe}\ \emph {et~al.}(2004)\citenamefont {McCabe},
  \citenamefont {Glotzer}, \citenamefont {Kieffer}, \citenamefont {Neurock},\
  and\ \citenamefont {Cummings}}]{mccabe2004multiscale}%
  \BibitemOpen
  \bibfield  {author} {\bibinfo {author} {\bibfnamefont {C.}~\bibnamefont
  {McCabe}}, \bibinfo {author} {\bibfnamefont {S.~C.}\ \bibnamefont {Glotzer}},
  \bibinfo {author} {\bibfnamefont {J.}~\bibnamefont {Kieffer}}, \bibinfo
  {author} {\bibfnamefont {M.}~\bibnamefont {Neurock}}, \ and\ \bibinfo
  {author} {\bibfnamefont {P.~T.}\ \bibnamefont {Cummings}},\ }\href@noop {}
  {\bibfield  {journal} {\bibinfo  {journal} {J. Comp. and Theor. Nanosci.}\
  }\textbf {\bibinfo {volume} {1}},\ \bibinfo {pages} {265} (\bibinfo {year}
  {2004})}\BibitemShut {NoStop}%
\bibitem [{\citenamefont {Liu}\ \emph {et~al.}(2011)\citenamefont {Liu},
  \citenamefont {Li}, \citenamefont {Lu}, \citenamefont {Zhang}, \citenamefont
  {Sun},\ and\ \citenamefont {Cui}}]{liu2011multiscale}%
  \BibitemOpen
  \bibfield  {author} {\bibinfo {author} {\bibfnamefont {H.}~\bibnamefont
  {Liu}}, \bibinfo {author} {\bibfnamefont {M.}~\bibnamefont {Li}}, \bibinfo
  {author} {\bibfnamefont {Z.-Y.}\ \bibnamefont {Lu}}, \bibinfo {author}
  {\bibfnamefont {Z.-G.}\ \bibnamefont {Zhang}}, \bibinfo {author}
  {\bibfnamefont {C.-C.}\ \bibnamefont {Sun}}, \ and\ \bibinfo {author}
  {\bibfnamefont {T.}~\bibnamefont {Cui}},\ }\href@noop {} {\bibfield
  {journal} {\bibinfo  {journal} {Macromol.}\ }\textbf {\bibinfo {volume}
  {44}},\ \bibinfo {pages} {8650} (\bibinfo {year} {2011})}\BibitemShut
  {NoStop}%
\bibitem [{\citenamefont {Peter}\ and\ \citenamefont
  {Kremer}(2009)}]{peter2009multiscale}%
  \BibitemOpen
  \bibfield  {author} {\bibinfo {author} {\bibfnamefont {C.}~\bibnamefont
  {Peter}}\ and\ \bibinfo {author} {\bibfnamefont {K.}~\bibnamefont {Kremer}},\
  }\href@noop {} {\bibfield  {journal} {\bibinfo  {journal} {Soft Matter}\
  }\textbf {\bibinfo {volume} {5}},\ \bibinfo {pages} {4357} (\bibinfo {year}
  {2009})}\BibitemShut {NoStop}%
\bibitem [{\citenamefont {Di~Pasquale}\ \emph {et~al.}(2012)\citenamefont
  {Di~Pasquale}, \citenamefont {Marchisio},\ and\ \citenamefont
  {Carbone}}]{di2012mixing}%
  \BibitemOpen
  \bibfield  {author} {\bibinfo {author} {\bibfnamefont {N.}~\bibnamefont
  {Di~Pasquale}}, \bibinfo {author} {\bibfnamefont {D.}~\bibnamefont
  {Marchisio}}, \ and\ \bibinfo {author} {\bibfnamefont {P.}~\bibnamefont
  {Carbone}},\ }\href@noop {} {\bibfield  {journal} {\bibinfo  {journal} {J.
  Chem. Phys.}\ }\textbf {\bibinfo {volume} {137}},\ \bibinfo {pages} {164111}
  (\bibinfo {year} {2012})}\BibitemShut {NoStop}%
\bibitem [{\citenamefont {Ensing}\ \emph {et~al.}(2007)\citenamefont {Ensing},
  \citenamefont {Nielsen}, \citenamefont {Moore}, \citenamefont {Klein},\ and\
  \citenamefont {Parrinello}}]{ensing2007energy}%
  \BibitemOpen
  \bibfield  {author} {\bibinfo {author} {\bibfnamefont {B.}~\bibnamefont
  {Ensing}}, \bibinfo {author} {\bibfnamefont {S.~O.}\ \bibnamefont {Nielsen}},
  \bibinfo {author} {\bibfnamefont {P.~B.}\ \bibnamefont {Moore}}, \bibinfo
  {author} {\bibfnamefont {M.~L.}\ \bibnamefont {Klein}}, \ and\ \bibinfo
  {author} {\bibfnamefont {M.}~\bibnamefont {Parrinello}},\ }\href@noop {}
  {\bibfield  {journal} {\bibinfo  {journal} {J. Chem. Theory and Comp.}\
  }\textbf {\bibinfo {volume} {3}},\ \bibinfo {pages} {1100} (\bibinfo {year}
  {2007})}\BibitemShut {NoStop}%
\bibitem [{\citenamefont {Lidorikis}\ \emph {et~al.}(2001)\citenamefont
  {Lidorikis}, \citenamefont {Bachlechner}, \citenamefont {Kalia},
  \citenamefont {Nakano}, \citenamefont {Vashishta},\ and\ \citenamefont
  {Voyiadjis}}]{lidorikis2001coupling}%
  \BibitemOpen
  \bibfield  {author} {\bibinfo {author} {\bibfnamefont {E.}~\bibnamefont
  {Lidorikis}}, \bibinfo {author} {\bibfnamefont {M.~E.}\ \bibnamefont
  {Bachlechner}}, \bibinfo {author} {\bibfnamefont {R.~K.}\ \bibnamefont
  {Kalia}}, \bibinfo {author} {\bibfnamefont {A.}~\bibnamefont {Nakano}},
  \bibinfo {author} {\bibfnamefont {P.}~\bibnamefont {Vashishta}}, \ and\
  \bibinfo {author} {\bibfnamefont {G.~Z.}\ \bibnamefont {Voyiadjis}},\
  }\href@noop {} {\bibfield  {journal} {\bibinfo  {journal} {Phys. Rev. Lett.}\
  }\textbf {\bibinfo {volume} {87}},\ \bibinfo {pages} {086104} (\bibinfo
  {year} {2001})}\BibitemShut {NoStop}%
\bibitem [{\citenamefont {Michel}\ \emph {et~al.}(2008)\citenamefont {Michel},
  \citenamefont {Orsi},\ and\ \citenamefont {Essex}}]{michel2008prediction}%
  \BibitemOpen
  \bibfield  {author} {\bibinfo {author} {\bibfnamefont {J.}~\bibnamefont
  {Michel}}, \bibinfo {author} {\bibfnamefont {M.}~\bibnamefont {Orsi}}, \ and\
  \bibinfo {author} {\bibfnamefont {J.~W.}\ \bibnamefont {Essex}},\ }\href@noop
  {} {\bibfield  {journal} {\bibinfo  {journal} {J. Phys. Chem. B}\ }\textbf
  {\bibinfo {volume} {112}},\ \bibinfo {pages} {657} (\bibinfo {year}
  {2008})}\BibitemShut {NoStop}%
\bibitem [{\citenamefont {Praprotnik}\ \emph {et~al.}(2008)\citenamefont
  {Praprotnik}, \citenamefont {Site},\ and\ \citenamefont
  {Kremer}}]{praprotnik2008multiscale}%
  \BibitemOpen
  \bibfield  {author} {\bibinfo {author} {\bibfnamefont {M.}~\bibnamefont
  {Praprotnik}}, \bibinfo {author} {\bibfnamefont {L.~D.}\ \bibnamefont
  {Site}}, \ and\ \bibinfo {author} {\bibfnamefont {K.}~\bibnamefont
  {Kremer}},\ }\href@noop {} {\bibfield  {journal} {\bibinfo  {journal} {Annu.
  Rev. Phys. Chem.}\ }\textbf {\bibinfo {volume} {59}},\ \bibinfo {pages} {545}
  (\bibinfo {year} {2008})}\BibitemShut {NoStop}%
\bibitem [{\citenamefont {Rzepiela}\ \emph {et~al.}(2011)\citenamefont
  {Rzepiela}, \citenamefont {Louhivuori}, \citenamefont {Peter},\ and\
  \citenamefont {Marrink}}]{rzepiela2011hybrid}%
  \BibitemOpen
  \bibfield  {author} {\bibinfo {author} {\bibfnamefont {A.~J.}\ \bibnamefont
  {Rzepiela}}, \bibinfo {author} {\bibfnamefont {M.}~\bibnamefont
  {Louhivuori}}, \bibinfo {author} {\bibfnamefont {C.}~\bibnamefont {Peter}}, \
  and\ \bibinfo {author} {\bibfnamefont {S.~J.}\ \bibnamefont {Marrink}},\
  }\href@noop {} {\bibfield  {journal} {\bibinfo  {journal} {Phys. Chem. Chem.
  Phys.}\ }\textbf {\bibinfo {volume} {13}},\ \bibinfo {pages} {10437}
  (\bibinfo {year} {2011})}\BibitemShut {NoStop}%
\bibitem [{\citenamefont {Werder}\ \emph {et~al.}(2005)\citenamefont {Werder},
  \citenamefont {Walther},\ and\ \citenamefont
  {Koumoutsakos}}]{werder2005hybrid}%
  \BibitemOpen
  \bibfield  {author} {\bibinfo {author} {\bibfnamefont {T.}~\bibnamefont
  {Werder}}, \bibinfo {author} {\bibfnamefont {J.~H.}\ \bibnamefont {Walther}},
  \ and\ \bibinfo {author} {\bibfnamefont {P.}~\bibnamefont {Koumoutsakos}},\
  }\href@noop {} {\bibfield  {journal} {\bibinfo  {journal} {J. Comp. Phys.}\
  }\textbf {\bibinfo {volume} {205}},\ \bibinfo {pages} {373} (\bibinfo {year}
  {2005})}\BibitemShut {NoStop}%
\bibitem [{\citenamefont {Heath}\ \emph {et~al.}(2007)\citenamefont {Heath},
  \citenamefont {Kavraki},\ and\ \citenamefont {Clementi}}]{heath2007coarse}%
  \BibitemOpen
  \bibfield  {author} {\bibinfo {author} {\bibfnamefont {A.~P.}\ \bibnamefont
  {Heath}}, \bibinfo {author} {\bibfnamefont {L.~E.}\ \bibnamefont {Kavraki}},
  \ and\ \bibinfo {author} {\bibfnamefont {C.}~\bibnamefont {Clementi}},\
  }\href@noop {} {\bibfield  {journal} {\bibinfo  {journal} {Proteins:
  Structure, Function, and Bioinformatics}\ }\textbf {\bibinfo {volume} {68}},\
  \bibinfo {pages} {646} (\bibinfo {year} {2007})}\BibitemShut {NoStop}%
\bibitem [{\citenamefont {Hess}\ \emph {et~al.}(2006)\citenamefont {Hess},
  \citenamefont {Holm},\ and\ \citenamefont {van~der Vegt}}]{hess2006osmotic}%
  \BibitemOpen
  \bibfield  {author} {\bibinfo {author} {\bibfnamefont {B.}~\bibnamefont
  {Hess}}, \bibinfo {author} {\bibfnamefont {C.}~\bibnamefont {Holm}}, \ and\
  \bibinfo {author} {\bibfnamefont {N.}~\bibnamefont {van~der Vegt}},\
  }\href@noop {} {\bibfield  {journal} {\bibinfo  {journal} {J. Chem. Phys.}\
  }\textbf {\bibinfo {volume} {124}},\ \bibinfo {pages} {164509} (\bibinfo
  {year} {2006})}\BibitemShut {NoStop}%
\bibitem [{\citenamefont {Liu}\ \emph {et~al.}(2008)\citenamefont {Liu},
  \citenamefont {Shi}, \citenamefont {Lyman},\ and\ \citenamefont
  {Voth}}]{liu2008reconstructing}%
  \BibitemOpen
  \bibfield  {author} {\bibinfo {author} {\bibfnamefont {P.}~\bibnamefont
  {Liu}}, \bibinfo {author} {\bibfnamefont {Q.}~\bibnamefont {Shi}}, \bibinfo
  {author} {\bibfnamefont {E.}~\bibnamefont {Lyman}}, \ and\ \bibinfo {author}
  {\bibfnamefont {G.~A.}\ \bibnamefont {Voth}},\ }\href@noop {} {\bibfield
  {journal} {\bibinfo  {journal} {J. Chem. Phys.}\ }\textbf {\bibinfo {volume}
  {129}},\ \bibinfo {pages} {114103} (\bibinfo {year} {2008})}\BibitemShut
  {NoStop}%
\bibitem [{\citenamefont {Rzepiela}\ \emph {et~al.}(2010)\citenamefont
  {Rzepiela}, \citenamefont {Sch{\"a}fer}, \citenamefont {Goga}, \citenamefont
  {Risselada}, \citenamefont {De~Vries},\ and\ \citenamefont
  {Marrink}}]{rzepiela2010reconstruction}%
  \BibitemOpen
  \bibfield  {author} {\bibinfo {author} {\bibfnamefont {A.~J.}\ \bibnamefont
  {Rzepiela}}, \bibinfo {author} {\bibfnamefont {L.~V.}\ \bibnamefont
  {Sch{\"a}fer}}, \bibinfo {author} {\bibfnamefont {N.}~\bibnamefont {Goga}},
  \bibinfo {author} {\bibfnamefont {H.~J.}\ \bibnamefont {Risselada}}, \bibinfo
  {author} {\bibfnamefont {A.~H.}\ \bibnamefont {De~Vries}}, \ and\ \bibinfo
  {author} {\bibfnamefont {S.~J.}\ \bibnamefont {Marrink}},\ }\href@noop {}
  {\bibfield  {journal} {\bibinfo  {journal} {J. Comp. Chem.}\ }\textbf
  {\bibinfo {volume} {31}},\ \bibinfo {pages} {1333} (\bibinfo {year}
  {2010})}\BibitemShut {NoStop}%
\bibitem [{\citenamefont {Reith}\ \emph {et~al.}(2003)\citenamefont {Reith},
  \citenamefont {P{\"u}tz},\ and\ \citenamefont
  {M{\"u}ller-Plathe}}]{reith2003deriving}%
  \BibitemOpen
  \bibfield  {author} {\bibinfo {author} {\bibfnamefont {D.}~\bibnamefont
  {Reith}}, \bibinfo {author} {\bibfnamefont {M.}~\bibnamefont {P{\"u}tz}}, \
  and\ \bibinfo {author} {\bibfnamefont {F.}~\bibnamefont
  {M{\"u}ller-Plathe}},\ }\href@noop {} {\bibfield  {journal} {\bibinfo
  {journal} {J. Comp. Chem.}\ }\textbf {\bibinfo {volume} {24}},\ \bibinfo
  {pages} {1624} (\bibinfo {year} {2003})}\BibitemShut {NoStop}%
\bibitem [{\citenamefont {Ercolessi}\ and\ \citenamefont
  {Adams}(1994)}]{ercolessi1994interatomic}%
  \BibitemOpen
  \bibfield  {author} {\bibinfo {author} {\bibfnamefont {F.}~\bibnamefont
  {Ercolessi}}\ and\ \bibinfo {author} {\bibfnamefont {J.~B.}\ \bibnamefont
  {Adams}},\ }\href@noop {} {\bibfield  {journal} {\bibinfo  {journal}
  {Europhys. Lett.}\ }\textbf {\bibinfo {volume} {26}},\ \bibinfo {pages} {583}
  (\bibinfo {year} {1994})}\BibitemShut {NoStop}%
\bibitem [{\citenamefont {Izvekov}\ and\ \citenamefont
  {Voth}(2005)}]{izvekov2005multiscale}%
  \BibitemOpen
  \bibfield  {author} {\bibinfo {author} {\bibfnamefont {S.}~\bibnamefont
  {Izvekov}}\ and\ \bibinfo {author} {\bibfnamefont {G.~A.}\ \bibnamefont
  {Voth}},\ }\href@noop {} {\bibfield  {journal} {\bibinfo  {journal} {J. Phys.
  Chem. B}\ }\textbf {\bibinfo {volume} {109}},\ \bibinfo {pages} {2469}
  (\bibinfo {year} {2005})}\BibitemShut {NoStop}%
\bibitem [{\citenamefont {Shell}(2008)}]{shell2008relative}%
  \BibitemOpen
  \bibfield  {author} {\bibinfo {author} {\bibfnamefont {M.~S.}\ \bibnamefont
  {Shell}},\ }\href@noop {} {\bibfield  {journal} {\bibinfo  {journal} {J.
  Chem. Phys.}\ }\textbf {\bibinfo {volume} {129}},\ \bibinfo {pages} {108}
  (\bibinfo {year} {2008})}\BibitemShut {NoStop}%
\bibitem [{\citenamefont {Chaimovich}\ and\ \citenamefont
  {Shell}(2011)}]{chaimovich2011coarse}%
  \BibitemOpen
  \bibfield  {author} {\bibinfo {author} {\bibfnamefont {A.}~\bibnamefont
  {Chaimovich}}\ and\ \bibinfo {author} {\bibfnamefont {M.~S.}\ \bibnamefont
  {Shell}},\ }\href@noop {} {\bibfield  {journal} {\bibinfo  {journal} {J.
  Chem. Phys.}\ }\textbf {\bibinfo {volume} {134}},\ \bibinfo {pages} {094112}
  (\bibinfo {year} {2011})}\BibitemShut {NoStop}%
\bibitem [{\citenamefont {Iacovella}\ \emph {et~al.}(2010)\citenamefont
  {Iacovella}, \citenamefont {Rogers}, \citenamefont {Glotzer},\ and\
  \citenamefont {Solomon}}]{iacovella2010pair}%
  \BibitemOpen
  \bibfield  {author} {\bibinfo {author} {\bibfnamefont {C.~R.}\ \bibnamefont
  {Iacovella}}, \bibinfo {author} {\bibfnamefont {R.~E.}\ \bibnamefont
  {Rogers}}, \bibinfo {author} {\bibfnamefont {S.~C.}\ \bibnamefont {Glotzer}},
  \ and\ \bibinfo {author} {\bibfnamefont {M.~J.}\ \bibnamefont {Solomon}},\
  }\href@noop {} {\bibfield  {journal} {\bibinfo  {journal} {J. Chem. Phys.}\
  }\textbf {\bibinfo {volume} {133}},\ \bibinfo {pages} {164903} (\bibinfo
  {year} {2010})}\BibitemShut {NoStop}%
\bibitem [{\citenamefont {Bozorgui}\ \emph {et~al.}(2013)\citenamefont
  {Bozorgui}, \citenamefont {Meng}, \citenamefont {Kumar}, \citenamefont
  {Chakravarty},\ and\ \citenamefont {Cacciuto}}]{bozorgui2013fluctuation}%
  \BibitemOpen
  \bibfield  {author} {\bibinfo {author} {\bibfnamefont {B.}~\bibnamefont
  {Bozorgui}}, \bibinfo {author} {\bibfnamefont {D.}~\bibnamefont {Meng}},
  \bibinfo {author} {\bibfnamefont {S.~K.}\ \bibnamefont {Kumar}}, \bibinfo
  {author} {\bibfnamefont {C.}~\bibnamefont {Chakravarty}}, \ and\ \bibinfo
  {author} {\bibfnamefont {A.}~\bibnamefont {Cacciuto}},\ }\href@noop {}
  {\bibfield  {journal} {\bibinfo  {journal} {Nano Lett.}\ }\textbf {\bibinfo
  {volume} {13}},\ \bibinfo {pages} {2732} (\bibinfo {year}
  {2013})}\BibitemShut {NoStop}%
\bibitem [{\citenamefont {Hadley}\ and\ \citenamefont
  {McCabe}(2010{\natexlab{a}})}]{hadley2010coarse}%
  \BibitemOpen
  \bibfield  {author} {\bibinfo {author} {\bibfnamefont {K.}~\bibnamefont
  {Hadley}}\ and\ \bibinfo {author} {\bibfnamefont {C.}~\bibnamefont
  {McCabe}},\ }\href@noop {} {\bibfield  {journal} {\bibinfo  {journal} {J.
  Chem. Phys.}\ }\textbf {\bibinfo {volume} {132}},\ \bibinfo {pages} {134505}
  (\bibinfo {year} {2010}{\natexlab{a}})}\BibitemShut {NoStop}%
\bibitem [{\citenamefont {Faller}(2004)}]{faller2004automatic}%
  \BibitemOpen
  \bibfield  {author} {\bibinfo {author} {\bibfnamefont {R.}~\bibnamefont
  {Faller}},\ }\href@noop {} {\bibfield  {journal} {\bibinfo  {journal}
  {Polymer}\ }\textbf {\bibinfo {volume} {45}},\ \bibinfo {pages} {3869}
  (\bibinfo {year} {2004})}\BibitemShut {NoStop}%
\bibitem [{\citenamefont {Fu}\ \emph {et~al.}(2012)\citenamefont {Fu},
  \citenamefont {Kulkarni}, \citenamefont {Shell},\ and\ \citenamefont
  {Leal}}]{fu2012test}%
  \BibitemOpen
  \bibfield  {author} {\bibinfo {author} {\bibfnamefont {C.-C.}\ \bibnamefont
  {Fu}}, \bibinfo {author} {\bibfnamefont {P.~M.}\ \bibnamefont {Kulkarni}},
  \bibinfo {author} {\bibfnamefont {M.~S.}\ \bibnamefont {Shell}}, \ and\
  \bibinfo {author} {\bibfnamefont {L.~G.}\ \bibnamefont {Leal}},\ }\href@noop
  {} {\bibfield  {journal} {\bibinfo  {journal} {J. Chem. Phys.}\ }\textbf
  {\bibinfo {volume} {137}},\ \bibinfo {pages} {164106} (\bibinfo {year}
  {2012})}\BibitemShut {NoStop}%
\bibitem [{\citenamefont {Qian}\ \emph {et~al.}(2008)\citenamefont {Qian},
  \citenamefont {Carbone}, \citenamefont {Chen}, \citenamefont
  {Karimi-Varzaneh}, \citenamefont {Liew},\ and\ \citenamefont
  {Müller-Plathe}}]{qian2008temperature}%
  \BibitemOpen
  \bibfield  {author} {\bibinfo {author} {\bibfnamefont {H.-J.}\ \bibnamefont
  {Qian}}, \bibinfo {author} {\bibfnamefont {P.}~\bibnamefont {Carbone}},
  \bibinfo {author} {\bibfnamefont {X.}~\bibnamefont {Chen}}, \bibinfo {author}
  {\bibfnamefont {H.~A.}\ \bibnamefont {Karimi-Varzaneh}}, \bibinfo {author}
  {\bibfnamefont {C.~C.}\ \bibnamefont {Liew}}, \ and\ \bibinfo {author}
  {\bibfnamefont {F.}~\bibnamefont {Müller-Plathe}},\ }\href@noop {}
  {\bibfield  {journal} {\bibinfo  {journal} {Macromol.}\ }\textbf {\bibinfo
  {volume} {41}},\ \bibinfo {pages} {9919} (\bibinfo {year}
  {2008})}\BibitemShut {NoStop}%
\bibitem [{\citenamefont {Bayramoglu}\ and\ \citenamefont
  {Faller}(2012)}]{bayramoglu2012coarse}%
  \BibitemOpen
  \bibfield  {author} {\bibinfo {author} {\bibfnamefont {B.}~\bibnamefont
  {Bayramoglu}}\ and\ \bibinfo {author} {\bibfnamefont {R.}~\bibnamefont
  {Faller}},\ }\href@noop {} {\bibfield  {journal} {\bibinfo  {journal}
  {Macromol.}\ }\textbf {\bibinfo {volume} {45}},\ \bibinfo {pages} {9205}
  (\bibinfo {year} {2012})}\BibitemShut {NoStop}%
\bibitem [{\citenamefont {Bayramoglu}\ and\ \citenamefont
  {Faller}(2013)}]{bayramoglu2013modeling}%
  \BibitemOpen
  \bibfield  {author} {\bibinfo {author} {\bibfnamefont {B.}~\bibnamefont
  {Bayramoglu}}\ and\ \bibinfo {author} {\bibfnamefont {R.}~\bibnamefont
  {Faller}},\ }\href@noop {} {\bibfield  {journal} {\bibinfo  {journal}
  {Macromol.}\ }\textbf {\bibinfo {volume} {46}},\ \bibinfo {pages} {7957}
  (\bibinfo {year} {2013})}\BibitemShut {NoStop}%
\bibitem [{\citenamefont {Ganguly}\ and\ \citenamefont {van~der
  Vegt}(2013)}]{ganguly2013convergence}%
  \BibitemOpen
  \bibfield  {author} {\bibinfo {author} {\bibfnamefont {P.}~\bibnamefont
  {Ganguly}}\ and\ \bibinfo {author} {\bibfnamefont {N.~F.}\ \bibnamefont
  {van~der Vegt}},\ }\href@noop {} {\bibfield  {journal} {\bibinfo  {journal}
  {J. Chem. Theory and Comp.}\ }\textbf {\bibinfo {volume} {9}},\ \bibinfo
  {pages} {1347} (\bibinfo {year} {2013})}\BibitemShut {NoStop}%
\bibitem [{\citenamefont {Chan}\ \emph {et~al.}(2007)\citenamefont {Chan},
  \citenamefont {Striolo}, \citenamefont {McCabe}, \citenamefont {Cummings},\
  and\ \citenamefont {Glotzer}}]{chan2007coarse}%
  \BibitemOpen
  \bibfield  {author} {\bibinfo {author} {\bibfnamefont {E.~R.}\ \bibnamefont
  {Chan}}, \bibinfo {author} {\bibfnamefont {A.}~\bibnamefont {Striolo}},
  \bibinfo {author} {\bibfnamefont {C.}~\bibnamefont {McCabe}}, \bibinfo
  {author} {\bibfnamefont {P.~T.}\ \bibnamefont {Cummings}}, \ and\ \bibinfo
  {author} {\bibfnamefont {S.~C.}\ \bibnamefont {Glotzer}},\ }\href@noop {}
  {\bibfield  {journal} {\bibinfo  {journal} {J. Chem. Phys.}\ }\textbf
  {\bibinfo {volume} {127}},\ \bibinfo {pages} {114102} (\bibinfo {year}
  {2007})}\BibitemShut {NoStop}%
\bibitem [{\citenamefont {Milano}\ \emph {et~al.}(2005)\citenamefont {Milano},
  \citenamefont {Goudeau},\ and\ \citenamefont
  {M{\"u}ller-Plathe}}]{milano2005multicentered}%
  \BibitemOpen
  \bibfield  {author} {\bibinfo {author} {\bibfnamefont {G.}~\bibnamefont
  {Milano}}, \bibinfo {author} {\bibfnamefont {S.}~\bibnamefont {Goudeau}}, \
  and\ \bibinfo {author} {\bibfnamefont {F.}~\bibnamefont
  {M{\"u}ller-Plathe}},\ }\href@noop {} {\bibfield  {journal} {\bibinfo
  {journal} {J. of Polymer Science Part B: Polymer Phys.}\ }\textbf {\bibinfo
  {volume} {43}},\ \bibinfo {pages} {871} (\bibinfo {year} {2005})}\BibitemShut
  {NoStop}%
\bibitem [{\citenamefont {Hadley}\ and\ \citenamefont
  {McCabe}(2010{\natexlab{b}})}]{hadley2010structurally}%
  \BibitemOpen
  \bibfield  {author} {\bibinfo {author} {\bibfnamefont {K.}~\bibnamefont
  {Hadley}}\ and\ \bibinfo {author} {\bibfnamefont {C.}~\bibnamefont
  {McCabe}},\ }\href@noop {} {\bibfield  {journal} {\bibinfo  {journal}
  {Biophys. J.}\ }\textbf {\bibinfo {volume} {99}},\ \bibinfo {pages} {2896}
  (\bibinfo {year} {2010}{\natexlab{b}})}\BibitemShut {NoStop}%
\bibitem [{\citenamefont {Rivera}\ \emph {et~al.}(2012)\citenamefont {Rivera},
  \citenamefont {Jennings},\ and\ \citenamefont
  {McCabe}}]{rivera2012examining}%
  \BibitemOpen
  \bibfield  {author} {\bibinfo {author} {\bibfnamefont {J.~L.}\ \bibnamefont
  {Rivera}}, \bibinfo {author} {\bibfnamefont {G.~K.}\ \bibnamefont
  {Jennings}}, \ and\ \bibinfo {author} {\bibfnamefont {C.}~\bibnamefont
  {McCabe}},\ }\href@noop {} {\bibfield  {journal} {\bibinfo  {journal} {J.
  Chem. Phys.}\ }\textbf {\bibinfo {volume} {136}},\ \bibinfo {pages} {244701}
  (\bibinfo {year} {2012})}\BibitemShut {NoStop}%
\bibitem [{Note1()}]{Note1}%
  \BibitemOpen
  \bibinfo {note} {\label {si}See supplemental material at [URL will be
  inserted by AIP] for derivation of $n$-dodecane pair and bonded potentials
  and associated RDFs}\BibitemShut {NoStop}%
\bibitem [{\citenamefont {Anderson}\ \emph {et~al.}(2008)\citenamefont
  {Anderson}, \citenamefont {Lorenz},\ and\ \citenamefont
  {Travesset}}]{anderson2008general}%
  \BibitemOpen
  \bibfield  {author} {\bibinfo {author} {\bibfnamefont {J.~A.}\ \bibnamefont
  {Anderson}}, \bibinfo {author} {\bibfnamefont {C.~D.}\ \bibnamefont
  {Lorenz}}, \ and\ \bibinfo {author} {\bibfnamefont {A.}~\bibnamefont
  {Travesset}},\ }\href@noop {} {\bibfield  {journal} {\bibinfo  {journal} {J.
  Comp. Phys.}\ }\textbf {\bibinfo {volume} {227}},\ \bibinfo {pages} {5342}
  (\bibinfo {year} {2008})}\BibitemShut {NoStop}%
\bibitem [{Note2()}]{Note2}%
  \BibitemOpen
  \bibinfo {note} {\lowercase
  {http://codeblue.umich.edu/hoomd-blue}}\BibitemShut {NoStop}%
\bibitem [{\citenamefont {Weeks}\ \emph {et~al.}(1971)\citenamefont {Weeks},
  \citenamefont {Chandler},\ and\ \citenamefont {Andersen}}]{weeks1971role}%
  \BibitemOpen
  \bibfield  {author} {\bibinfo {author} {\bibfnamefont {J.~D.}\ \bibnamefont
  {Weeks}}, \bibinfo {author} {\bibfnamefont {D.}~\bibnamefont {Chandler}}, \
  and\ \bibinfo {author} {\bibfnamefont {H.~C.}\ \bibnamefont {Andersen}},\
  }\href {\doibase http://dx.doi.org/10.1063/1.1674820} {\bibfield  {journal}
  {\bibinfo  {journal} {J. Chem. Phys.}\ }\textbf {\bibinfo {volume} {54}},\
  \bibinfo {pages} {5237} (\bibinfo {year} {1971})}\BibitemShut {NoStop}%
\bibitem [{\citenamefont {Pu}\ \emph {et~al.}(2007)\citenamefont {Pu},
  \citenamefont {Leng}, \citenamefont {Zhao},\ and\ \citenamefont
  {Cummings}}]{pu2007molecular}%
  \BibitemOpen
  \bibfield  {author} {\bibinfo {author} {\bibfnamefont {Q.}~\bibnamefont
  {Pu}}, \bibinfo {author} {\bibfnamefont {Y.}~\bibnamefont {Leng}}, \bibinfo
  {author} {\bibfnamefont {X.}~\bibnamefont {Zhao}}, \ and\ \bibinfo {author}
  {\bibfnamefont {P.~T.}\ \bibnamefont {Cummings}},\ }\href@noop {} {\bibfield
  {journal} {\bibinfo  {journal} {Nanotech.}\ }\textbf {\bibinfo {volume}
  {18}},\ \bibinfo {pages} {424007} (\bibinfo {year} {2007})}\BibitemShut
  {NoStop}%
\bibitem [{\citenamefont {Keys}\ \emph {et~al.}(2011)\citenamefont {Keys},
  \citenamefont {Iacovella},\ and\ \citenamefont
  {Glotzer}}]{keys2011characterizing}%
  \BibitemOpen
  \bibfield  {author} {\bibinfo {author} {\bibfnamefont {A.~S.}\ \bibnamefont
  {Keys}}, \bibinfo {author} {\bibfnamefont {C.~R.}\ \bibnamefont {Iacovella}},
  \ and\ \bibinfo {author} {\bibfnamefont {S.~C.}\ \bibnamefont {Glotzer}},\
  }\href@noop {} {\bibfield  {journal} {\bibinfo  {journal} {J. Comp. Phys.}\
  }\textbf {\bibinfo {volume} {230}},\ \bibinfo {pages} {6438} (\bibinfo {year}
  {2011})}\BibitemShut {NoStop}%
\bibitem [{\citenamefont {Heller}\ \emph {et~al.}(1993)\citenamefont {Heller},
  \citenamefont {Schaefer},\ and\ \citenamefont
  {Schulten}}]{heller1993molecular}%
  \BibitemOpen
  \bibfield  {author} {\bibinfo {author} {\bibfnamefont {H.}~\bibnamefont
  {Heller}}, \bibinfo {author} {\bibfnamefont {M.}~\bibnamefont {Schaefer}}, \
  and\ \bibinfo {author} {\bibfnamefont {K.}~\bibnamefont {Schulten}},\
  }\href@noop {} {\bibfield  {journal} {\bibinfo  {journal} {J. Phys. Chem.}\
  }\textbf {\bibinfo {volume} {97}},\ \bibinfo {pages} {8343} (\bibinfo {year}
  {1993})}\BibitemShut {NoStop}%
\bibitem [{\citenamefont {Egberts}\ \emph {et~al.}(1994)\citenamefont
  {Egberts}, \citenamefont {Marrink},\ and\ \citenamefont
  {Berendsen}}]{egberts1994molecular}%
  \BibitemOpen
  \bibfield  {author} {\bibinfo {author} {\bibfnamefont {E.}~\bibnamefont
  {Egberts}}, \bibinfo {author} {\bibfnamefont {S.-J.}\ \bibnamefont
  {Marrink}}, \ and\ \bibinfo {author} {\bibfnamefont {H.~J.}\ \bibnamefont
  {Berendsen}},\ }\href@noop {} {\bibfield  {journal} {\bibinfo  {journal}
  {Euro. Biophys. J.}\ }\textbf {\bibinfo {volume} {22}},\ \bibinfo {pages}
  {423} (\bibinfo {year} {1994})}\BibitemShut {NoStop}%
\bibitem [{\citenamefont {Saiz}\ and\ \citenamefont
  {Klein}(2002)}]{saiz2002computer}%
  \BibitemOpen
  \bibfield  {author} {\bibinfo {author} {\bibfnamefont {L.}~\bibnamefont
  {Saiz}}\ and\ \bibinfo {author} {\bibfnamefont {M.~L.}\ \bibnamefont
  {Klein}},\ }\href@noop {} {\bibfield  {journal} {\bibinfo  {journal}
  {Accounts of Chemical Research}\ }\textbf {\bibinfo {volume} {35}},\ \bibinfo
  {pages} {482} (\bibinfo {year} {2002})}\BibitemShut {NoStop}%
\bibitem [{\citenamefont {Scott}(2002)}]{scott2002modeling}%
  \BibitemOpen
  \bibfield  {author} {\bibinfo {author} {\bibfnamefont {H.~L.}\ \bibnamefont
  {Scott}},\ }\href@noop {} {\bibfield  {journal} {\bibinfo  {journal} {Current
  Opinion in Structural Biology}\ }\textbf {\bibinfo {volume} {12}},\ \bibinfo
  {pages} {495} (\bibinfo {year} {2002})}\BibitemShut {NoStop}%
\bibitem [{\citenamefont {Venturoli}\ \emph {et~al.}(2006)\citenamefont
  {Venturoli}, \citenamefont {Maddalena~Sperotto}, \citenamefont {Kranenburg},\
  and\ \citenamefont {Smit}}]{venturoli2006mesoscopic}%
  \BibitemOpen
  \bibfield  {author} {\bibinfo {author} {\bibfnamefont {M.}~\bibnamefont
  {Venturoli}}, \bibinfo {author} {\bibfnamefont {M.}~\bibnamefont
  {Maddalena~Sperotto}}, \bibinfo {author} {\bibfnamefont {M.}~\bibnamefont
  {Kranenburg}}, \ and\ \bibinfo {author} {\bibfnamefont {B.}~\bibnamefont
  {Smit}},\ }\href@noop {} {\bibfield  {journal} {\bibinfo  {journal} {Physics
  Reports}\ }\textbf {\bibinfo {volume} {437}},\ \bibinfo {pages} {1} (\bibinfo
  {year} {2006})}\BibitemShut {NoStop}%
\bibitem [{\citenamefont {Vettorel}\ and\ \citenamefont
  {Meyer}(2006)}]{vettorel2006coarse}%
  \BibitemOpen
  \bibfield  {author} {\bibinfo {author} {\bibfnamefont {T.}~\bibnamefont
  {Vettorel}}\ and\ \bibinfo {author} {\bibfnamefont {H.}~\bibnamefont
  {Meyer}},\ }\href@noop {} {\bibfield  {journal} {\bibinfo  {journal} {J.
  Chem. Theory and Comp.}\ }\textbf {\bibinfo {volume} {2}},\ \bibinfo {pages}
  {616} (\bibinfo {year} {2006})}\BibitemShut {NoStop}%
\bibitem [{\citenamefont {D'Adamo}\ \emph {et~al.}(2012)\citenamefont
  {D'Adamo}, \citenamefont {Pelissetto},\ and\ \citenamefont
  {Pierleoni}}]{d2012coarse}%
  \BibitemOpen
  \bibfield  {author} {\bibinfo {author} {\bibfnamefont {G.}~\bibnamefont
  {D'Adamo}}, \bibinfo {author} {\bibfnamefont {A.}~\bibnamefont {Pelissetto}},
  \ and\ \bibinfo {author} {\bibfnamefont {C.}~\bibnamefont {Pierleoni}},\
  }\href@noop {} {\bibfield  {journal} {\bibinfo  {journal} {Soft Matter}\
  }\textbf {\bibinfo {volume} {8}},\ \bibinfo {pages} {5151} (\bibinfo {year}
  {2012})}\BibitemShut {NoStop}%
\bibitem [{\citenamefont {Hess}\ \emph {et~al.}(2008)\citenamefont {Hess},
  \citenamefont {Kutzner}, \citenamefont {Van Der~Spoel},\ and\ \citenamefont
  {Lindahl}}]{hess2008gromacs}%
  \BibitemOpen
  \bibfield  {author} {\bibinfo {author} {\bibfnamefont {B.}~\bibnamefont
  {Hess}}, \bibinfo {author} {\bibfnamefont {C.}~\bibnamefont {Kutzner}},
  \bibinfo {author} {\bibfnamefont {D.}~\bibnamefont {Van Der~Spoel}}, \ and\
  \bibinfo {author} {\bibfnamefont {E.}~\bibnamefont {Lindahl}},\ }\href@noop
  {} {\bibfield  {journal} {\bibinfo  {journal} {J. Chem. Theory and Comp.}\
  }\textbf {\bibinfo {volume} {4}},\ \bibinfo {pages} {435} (\bibinfo {year}
  {2008})}\BibitemShut {NoStop}%
\bibitem [{\citenamefont {Plimpton}(1995)}]{plimpton1995fast}%
  \BibitemOpen
  \bibfield  {author} {\bibinfo {author} {\bibfnamefont {S.}~\bibnamefont
  {Plimpton}},\ }\href@noop {} {\bibfield  {journal} {\bibinfo  {journal} {J.
  Comp. Phys.}\ }\textbf {\bibinfo {volume} {117}},\ \bibinfo {pages} {1}
  (\bibinfo {year} {1995})}\BibitemShut {NoStop}%
\bibitem [{\citenamefont {Brown}\ \emph {et~al.}(2011)\citenamefont {Brown},
  \citenamefont {Wang}, \citenamefont {Plimpton},\ and\ \citenamefont
  {Tharrington}}]{brown2011implementing}%
  \BibitemOpen
  \bibfield  {author} {\bibinfo {author} {\bibfnamefont {W.~M.}\ \bibnamefont
  {Brown}}, \bibinfo {author} {\bibfnamefont {P.}~\bibnamefont {Wang}},
  \bibinfo {author} {\bibfnamefont {S.~J.}\ \bibnamefont {Plimpton}}, \ and\
  \bibinfo {author} {\bibfnamefont {A.~N.}\ \bibnamefont {Tharrington}},\
  }\href@noop {} {\bibfield  {journal} {\bibinfo  {journal} {Computer Phys.
  Comm.}\ }\textbf {\bibinfo {volume} {182}},\ \bibinfo {pages} {898} (\bibinfo
  {year} {2011})}\BibitemShut {NoStop}%
\bibitem [{\citenamefont {Brown}\ \emph {et~al.}(2012)\citenamefont {Brown},
  \citenamefont {Kohlmeyer}, \citenamefont {Plimpton},\ and\ \citenamefont
  {Tharrington}}]{brown2012implementing}%
  \BibitemOpen
  \bibfield  {author} {\bibinfo {author} {\bibfnamefont {W.~M.}\ \bibnamefont
  {Brown}}, \bibinfo {author} {\bibfnamefont {A.}~\bibnamefont {Kohlmeyer}},
  \bibinfo {author} {\bibfnamefont {S.~J.}\ \bibnamefont {Plimpton}}, \ and\
  \bibinfo {author} {\bibfnamefont {A.~N.}\ \bibnamefont {Tharrington}},\
  }\href@noop {} {\bibfield  {journal} {\bibinfo  {journal} {Computer Phys.
  Comm.}\ }\textbf {\bibinfo {volume} {183}},\ \bibinfo {pages} {449} (\bibinfo
  {year} {2012})}\BibitemShut {NoStop}%
\end{thebibliography}%
\end{document}